%% file: main.tex
\documentclass[format=acmsmall, review=false, screen=true]{acmart}

\usepackage{booktabs} 

\usepackage[ruled]{algorithm2e} 

\SetAlFnt{\small}
\SetAlCapFnt{\small}
\SetAlCapNameFnt{\small}
\SetAlCapHSkip{0pt}
\IncMargin{-\parindent}

\usepackage{graphicx}
\usepackage{balance}
\usepackage{comment}
\usepackage{subfigure}
\usepackage{booktabs}
\usepackage{amsmath}
\usepackage{amssymb}
\usepackage{amsfonts}
\usepackage{bm}
\usepackage{colortbl}
\usepackage{tabularx}
\usepackage{color}
\usepackage{epsfig}
\usepackage{xspace}
\usepackage{hyperref}    
\hypersetup{pdfstartview=FitH}
\usepackage{graphicx}    
\usepackage{url}         %
\usepackage{algpseudocode}
\usepackage{placeins}
\usepackage{multirow}
\usepackage{epstopdf}
\usepackage{tabularx}
\usepackage[colorinlistoftodos]{todonotes}
\usepackage{blindtext}

\received{February 2007}
\received[revised]{March 2009}
\received[accepted]{June 2009}

\begin{document}
\title[]{No Need of Data Pre-processing: A General Framework for Radio-Based Device-Free Context Awareness}

\author{Bo Wei}
\affiliation{%
  \institution{Northumbria University}
  \city{Newcastle upon Tyne}
  \state{Tyne and Wear}
  \country{UK}}
\email{bo.wei@northumbria.ac.uk}

\author{Kai Li}
\affiliation{%
  \institution{Research Centre in Real-Time and Embedded Computing System (CISTER)}
  \city{Porto}
  \country{Portugal}
}
\email{kaili@isep.ipp.pt}

\author{Chengwen Luo}
\affiliation{%
 \institution{Shenzhen University}
 \city{Shenzhen}
 \country{China}}
\email{chengwen@szu.edu.cn}

\author{Weitao Xu}
\affiliation{%
  \institution{City University of Hong Kong}
  \city{Hong Kong}
  \country{China}
}
\email{weitao.xu@cityu.edu.hk}

\author{Jin Zhang}
\affiliation{%
  \institution{Chinese Academy of Sciences}
  \city{Shenzhen}
  \country{China}
}
\email{jin.zhang@siat.ac.cn}

\begin{abstract}
Device-free context awareness is important to many applications. There are two broadly used approaches for device-free context awareness, i.e. video-based and radio-based. Video-based applications can deliver good performance, but privacy is a serious concern. Radio-based context awareness has drawn researchers' attention instead because it does not violate privacy and radio signal can penetrate obstacles. Recently, deep learning has been introduced into radio-based device-free context awareness and helps boost the recognition accuracy. The present works design explicit methods for each radio based application.  They also use one additional step to extract features before conducting classification and exploit deep learning as a classification tool. The additional initial data processing step introduces unnecessary noise and information loss. Without initial data processing, it is, however, challenging to explore patterns of raw signals. In this paper, we are the first to propose an innovative deep learning based general framework for both signal processing and classification. The key novelty of this paper is that the framework can be generalised for all the radio-based context awareness applications. We also eliminate the additional effort to extract features from raw radio signals. We conduct extensive evaluations to show the superior performance of our proposed method and its generalisation. 

\end{abstract}

%
%

\ccsdesc[500]{Networks~Sensor networks}
\ccsdesc[500]{Computer systems organization~Sensor networks}

%
%

\keywords{Device-free, channel state information, deep learning, context awareness}

\maketitle

\input{Tex/intro}

\input{Tex/related}
\input{Tex/background}
\input{Tex/method}
\input{Tex/experiment}

\section{Conclusion}\label{sec:conclusion}
In this paper, we propose an innovative DNN based framework for radio-based device-free context awareness applications. We are the first to design a general framework of radio-based based applications, and we show the generalisation using extensive experiments.  
Without any data pre-processing, our proposed model can still achieve equivalent or better performance.
We use public radio-based gesture recognition datasets to evaluate our proposed model, which results in approximately 100\% accuracy for recognising 276 gestures. We also show the capability of our proposed framework for radio-based applications in RFI environments. We also discuss the design of DNN structures using our proposed general framework and show the effects of different layers in the DNN models in both datasets.

\bibliographystyle{ACM-Reference-Format}
\bibliography{sigproc_short}

\end{document}

%% file: Tex/intro.tex

\section{Introduction}\label{intro}
\label{sec:intro}

Context awareness, e.g. activity recognition, gesture recognition, etc., is a popular research direction and an essential component in many fields, such as health monitoring, emergency response, etc. Many approaches based on cameras and wearable sensors have been designed \cite{kinect,xiaomiwristband,jawboneup}.
Camera-based solutions use informative images and show excellent performance for context recognition, especially when integrating additional depth sensors, e.g. Microsoft Kinect \cite{kinect}. 
Motion-based context awareness applications are also prevalent by taking advantage of widely used motion sensors in mobile phones and smartwatches. 
Unfortunately, these solutions still have their limitations. 
Camera-based solutions have a privacy concern, especially when the captured images have to be stored and processed in the cloud. There is no privacy concern for motion sensor-based approaches, but wearable devices must be carried and worn in proper manners to collect data by built-in sensors. This is inconvenient for older adults who may often forget to carry wearable devices. 
The wireless radio signal has broadly been investigated for \emph{device-free} context awareness to resolve these issues. 
Halperin et al. \cite{halperin2011tool} introduced a custom modified firmware based on a standard off-the-shelf WiFi card, Intel 5300 802.11n MIMO radio, which can provide informative Channel State Information (CSI) from Physical Layer. 
%
%
CSI measurements significantly increase resolution compared with the traditional Received Signal Strength (RSS), which helps realise many radio-based device free context awareness applications. 
The WiFi devices are widely deployed in an indoor environment, such as office and home, etc., and its ubiquity further stimulates its large-scale device-free application.
CSI measurements collected from multiple radio devices deployed in the area of interests (AoI) can capture radio interference caused by monitored people.
The patterns from CSI measurements due to the multi-path effect are mapped to relevant contexts. 
Recently, emerging radio-based context awareness applications (such as activity recognition \cite{WeiActivity:2015,wei2019from,wang2014eyes}, gesture recognition \cite{ma2018signfi}, ``lip-read'' \cite{wang2014we}, identity recognition \cite{zhang2016wifi}, etc) have been studied and can have equivalent performance as the traditional applications using wearable devices and cameras. 
The radio-based device-free context awareness applications gain a competitive advantage without requiring people to carry any device and breaking the privacy. 


 
CSI based context awareness applications widely use supervised machine learning techniques which require the training stage to analyse labelled CSI measurements and derive a context awareness model. Both classic machine learning techniques ( e.g., k-Nearest Neighbour \cite{sen2012you}, Sparse Representation Classification\cite{Wei:2015gx,zhang2016wifi}, etc.) and deep learning techniques (e.g. \cite{ma2018signfi,wang2018spatial,wang2015deepfi, wang2017csi}) have also been introduced for CSI based applications. 
One customised method is particularly designed for each application. An initial step for CSI data pre-processing is further used to extract features for classification. 
Raw complex-valued CSI measurements have been considered as no usable and recognisable patterns without any data pre-processing due to their asynchronous phase. The current research works all use the initial data processing as the first step.
As a result, recognition performance is highly relevant to these data processing methods. 
In contrast, the other successful fields of state-of-the-art machine learning applications, such as computer vision, use an entirely different strategy for pattern recognition. 
Without using any initial data processing methods, most computer vision techniques use raw images to train deep neural network (DNN) models and still significantly outperform traditional methods. They also use a general framework for computer vision-based applications \cite{simonyan2014very}. They only need to modify the configurations or architectures for each particular application, but the whole DNN structure is still based on that general framework. Motivated by this, we design a novel DNN based general framework using raw CSI measurements without any additional feature selection and data pre-processing. We aim to generalise our framework to radio-based device-free context awareness applications.  Our proposed method can preserve all the information without any pattern loss caused by an initial data processing stage, which equips itself the ability of the generalisation. Capable of both signal processing and classification, our proposed DNN framework attains excellent performance. 
As far as we know, we are the first to propose a general framework for radio-based application and directly use the \emph{raw} CSI measurements to train the DNN model without any pre-processing.


To summarise, the contributions of this paper are as follows:

\begin{itemize}

\item To the best of our knowledge, we are the first to propose an innovative deep learning  based \emph{framework} using raw CSI data for context awareness without any feature selection and initial data pre-processing. This model is capable of being generalised to any radio-based context awareness applications. 

\item We show the intuition and theoretical explanations in detail of applying the raw CSI on training the DNN model. We also reveal the reason why the proposed general framework from raw CSI measurements can improve performance.

\item We conduct extensive evaluations on various datasets to demonstrate the generalisation of our framework.  We first design the DNN structure based on the general framework for radio-based device-free gesture recognition. We evaluate our proposed structure by using public CSI gesture recognition datasets that include 270 gestures and achieve approximately 100\% accuracy.  

\item We further take advantage of our proposed general framework to design a DNN structure for activity recognition. We perform evaluations on CSI based activity recognition datasets in radio frequency interfered environment, which achieves equivalent performance as the previous work without any data processing. 

\end{itemize}

The rest of the paper is organised as follows. Section \ref{sec:related} is related work. Section \ref{sec:background} shows the background of CSI and the motivation of our method. The proposed DNN model is illustrated in Section \ref{sec:methods} and evaluated in Section \ref{sec:experiments}. Section \ref{sec:conclusion} concludes the paper.

%% file: Tex/related.tex
\section{Related work}
\label{sec:related}
In this section, we discuss related research works in context awareness from radio data. 
The objects and people in one environment can reflect and diffract radio waves due to the multiple-path effect. This radio frequency perturbations can be detected by both the coarse-grained and fine-grained radio signal characteristics. 

Software-defined radio has been used for context awareness applications. 
Radio signals have also been used to perform gesture recognition. WiSee designed by Adib et al.~\cite{Adib:2013:STW:2486001.2486039} and WiVi designed by Pu et al.~\cite{pu2013whole} used software-defined radio to extract the Doppler effect caused by the gesture. 
In order to reduce energy consumption, Kellogg et al.~\cite{Kellogg:2014} built AllSee that uses RFID tags and power-harvesting sensors for gesture recognition.
To further improve the resolution of the radio signal based localisation and gesture recognition, Adib et al.~\cite{Adib:2014} designed WiTrack and obtained time-of-flight from the Frequency Modulated Carrier Wave~(FMCW) technology for localisation in 3 dimensions. 

Instead of using special software-defined radio equipment for those applications, RSS, which can be readily obtained from off-the-shelf radio cards, have been broadly used in device-free localisation applications\cite{WilsonRTI:2010,ZhaoNoise2011,KaltiokallioBP12,ZhaoKRTI:2013,WeidRTI:2015,Xu:2012:sensys}. 
The coarse RSS was also successfully applied for gesture recognition \cite{melgarejo2014leveraging}. 

However, sufficient characteristic information cannot be acquired from RSSI. To fulfil the gap, Halperin et al. \cite{Halperin_csitool} and Xie et al.~\cite{xie2015precise} modified drivers of off-the-shelf WiFi cards, i.e. Intel 5300 and Atheros 9390, and obtain fine-grained Channel State Information (CSI) from the Physical layer of the customised hardware. The acquaintance of fine-grained features motivates many radio-based applications. 
Sen et al. used modified WiFi 5300 cards and collect CSI for indoor localisation \cite{sen2012you}. Zhou et al. \cite{zhou2013omnidirectional} also applied CSI information for omnidirectional passive human detection. Gesture recognition application and activity recognition applications have been proposed by using fine-grained CSI information\cite{melgarejo2014leveraging,ma2018signfi,wang2018spatial,Wei:2015gx,wang2014eyes}. 
CSI based lip-read, emotion recognition applications and identity recognition applications were designed in \cite{wang2014we,zhao2016emotion,zeng2016wiwho,zhang2016wifi,wang2016gait}. 
There are also other context awareness research works using CSI, such as breath detection\cite{zhang2019breathtrack}, sleep monitoring\cite{liu2018monitoring} and motion detection\cite{gu2017mosense}. CSI is also used for multiuser MIMO transmissions\cite{nam2015deltasnr}.
Moreover, phase information from CSI is used by research works for radio-based device-free applications as additional supplementary information \cite{qian2017inferring,wang2017csi,wang2017phasebeat,ma2018signfi}.    

Deep learning, a branch of machine learning, is introduced for radio-based context awareness applications. 
Wang et al. \cite{wang2015deepfi, wang2017csi} designed indoor localisation systems by using deep learning techniques along with CSI phase information. \cite{wang2018spatial} et al. also designed a deep learning model for CSI based activity recognition.  Ma et al. \cite{ma2018signfi} took advantage of deep learning techniques and showed the feasibility of using 20 MHz radio band to recognise 276 gestures. These research works process CSI data initially and use deep learning as a classification tool. A specialised method is designed for each application. Different from these works, we propose a general framework for radio-based device-free applications.  Our proposed method conducts context awareness with raw CSI and outperforms the existing work without requiring any pre-processed CSI. 

%% file: Tex/background.tex
\begin{figure*}[!ht]
    \centering
    \subfigure[Raw data in the complex plane]{
         \includegraphics[scale = 0.25]{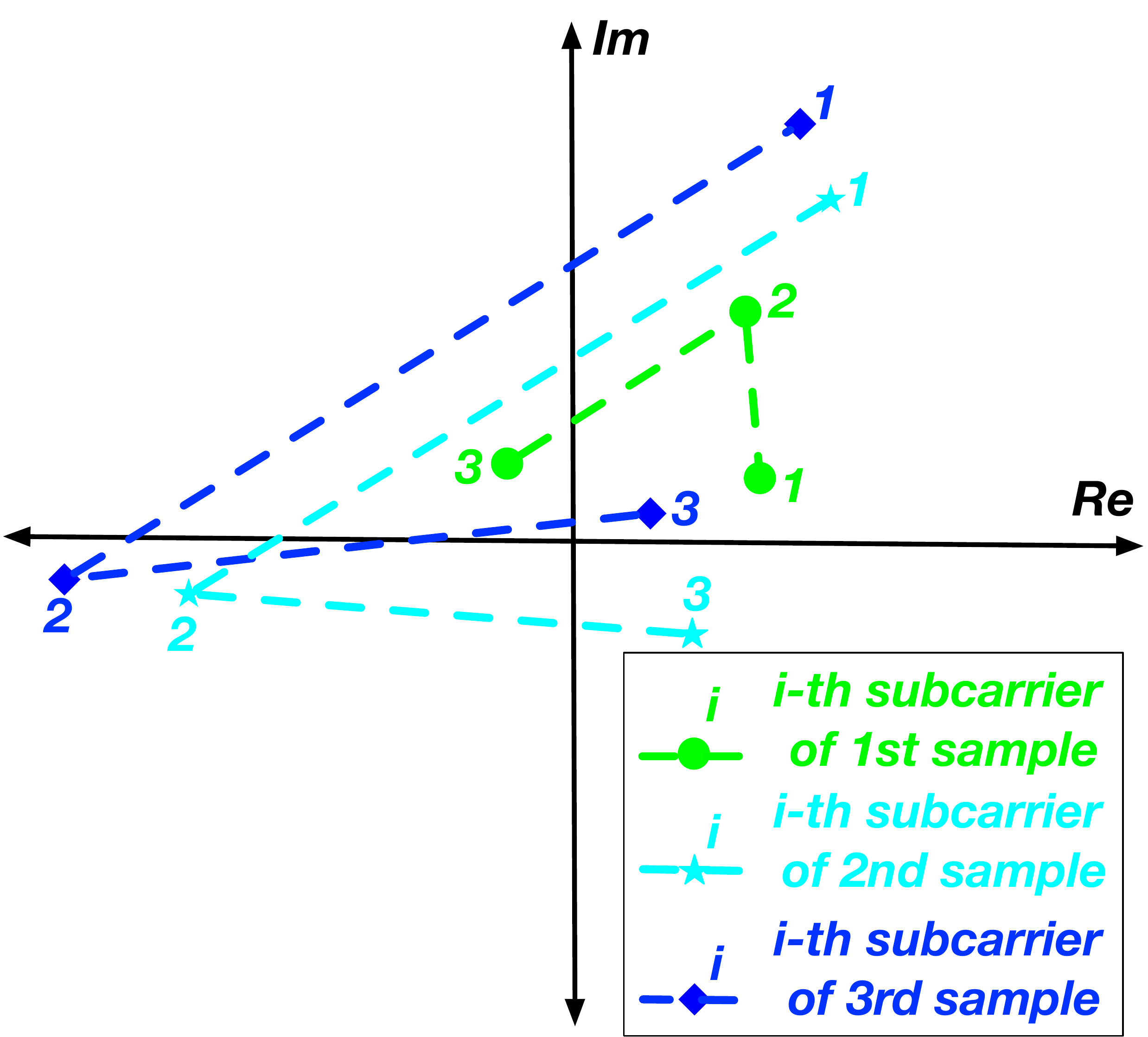}
        \label{fig:complex_plane}
    }
    \subfigure[Raw amplitude]{
         \includegraphics[scale = 0.25]{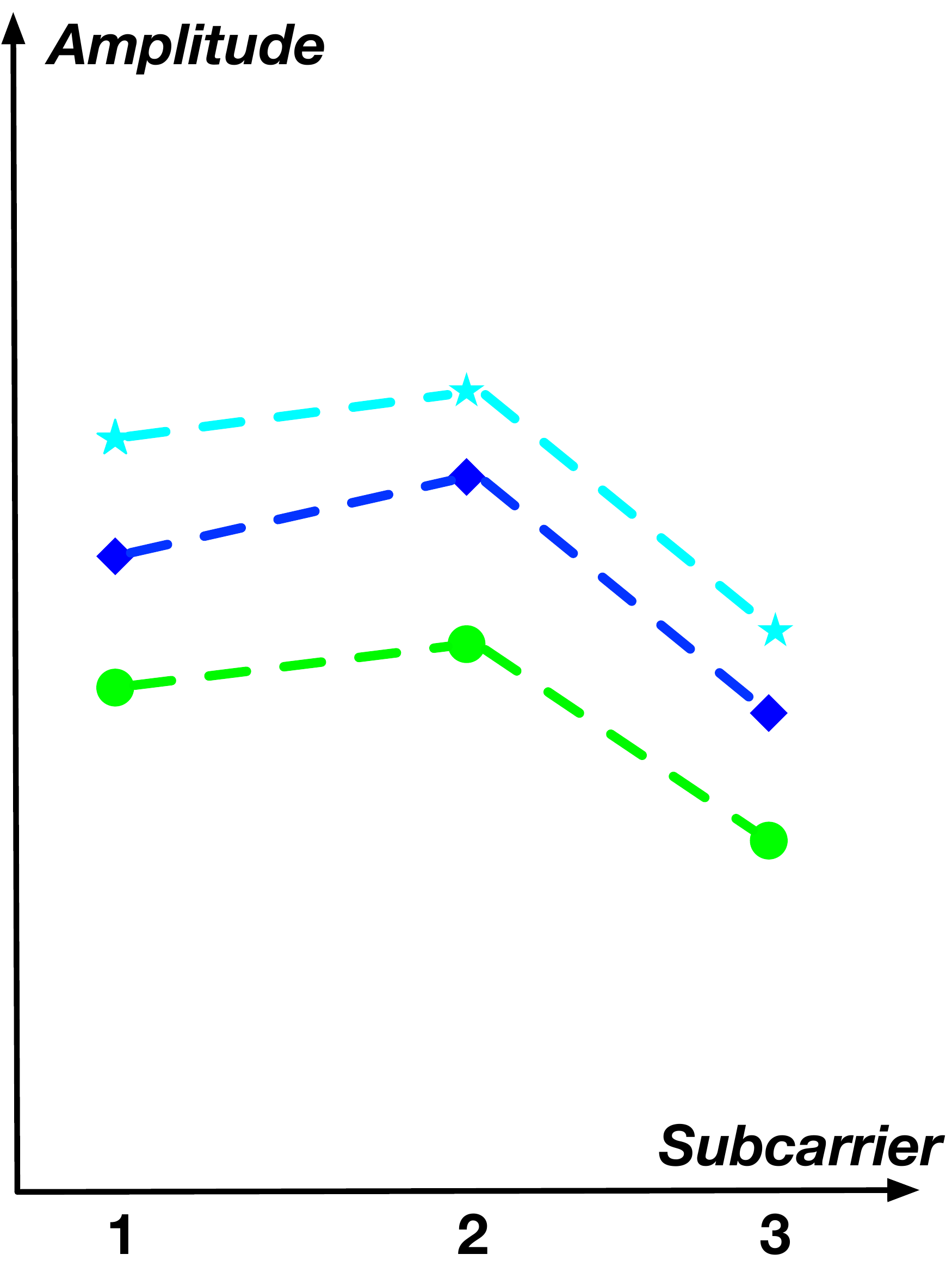}
        \label{fig:amplitude}
    }

    \caption{Examples of data processing - Part 1}
   \label{fig:3samples-p1}
\end{figure*}

\begin{figure*}[!ht]
    \centering

    \subfigure[Normalised amplitude]{
         \includegraphics[scale = 0.25]{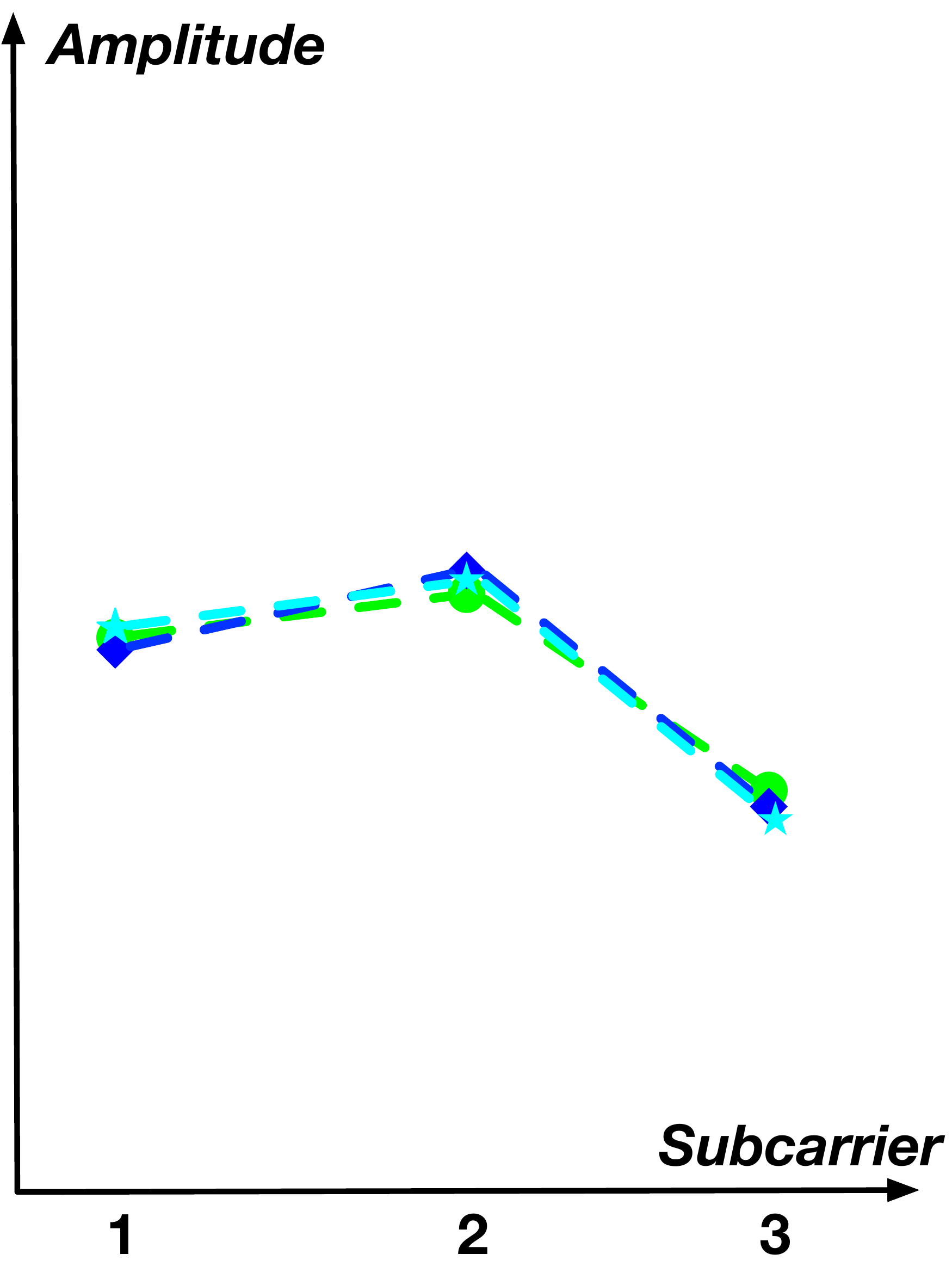}
        \label{fig:amplitude_normalised}
    }
    \subfigure[Raw phase]{
         \includegraphics[scale = 0.25]{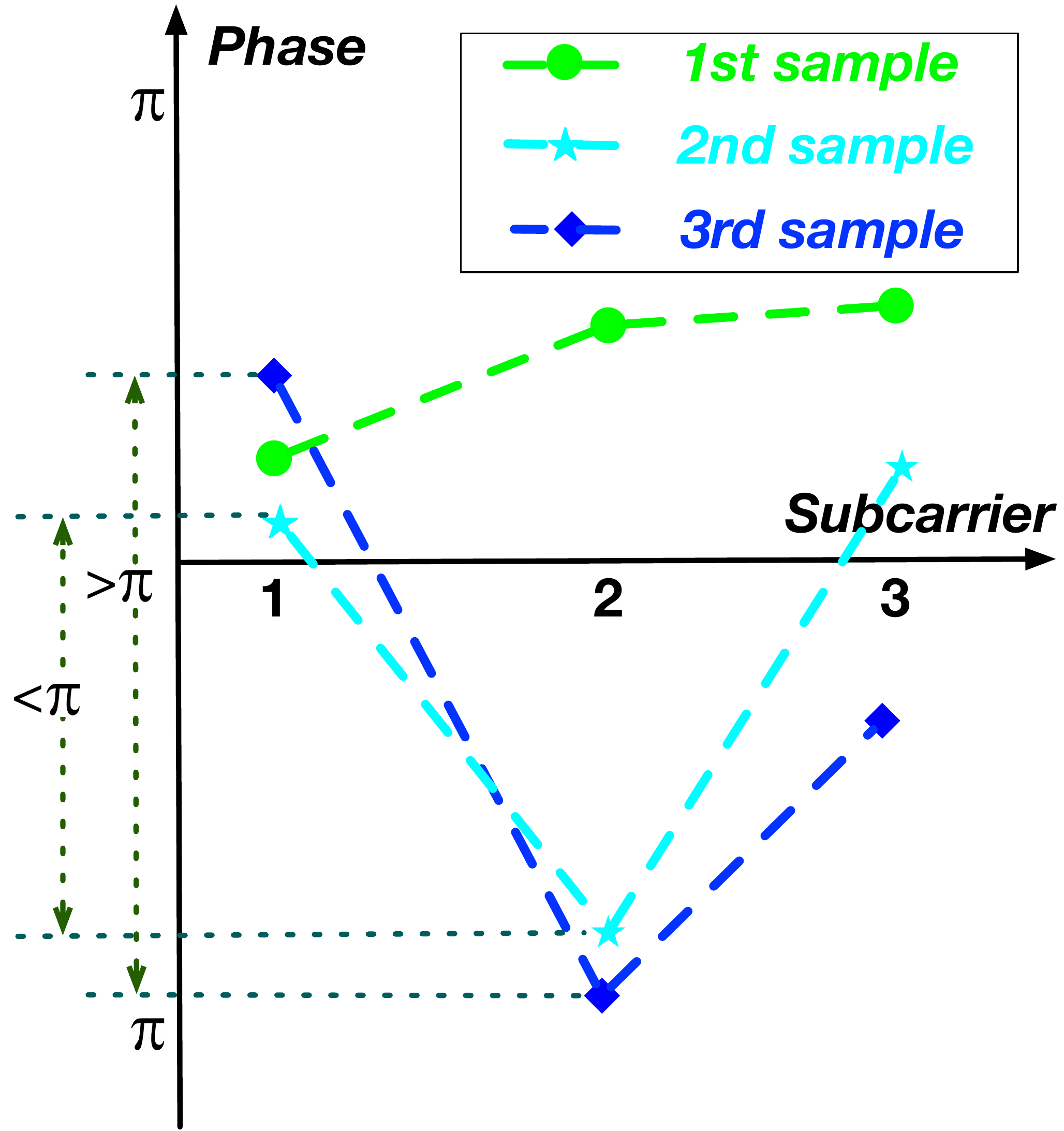}
        \label{fig:phase_raw}
    }

    \caption{Examples of data processing - Part 2}
   \label{fig:3samples-p2}
\end{figure*}

\begin{figure*}[!ht]
    \centering
    \subfigure[Phase after unwrapping]{
         \includegraphics[scale = 0.25]{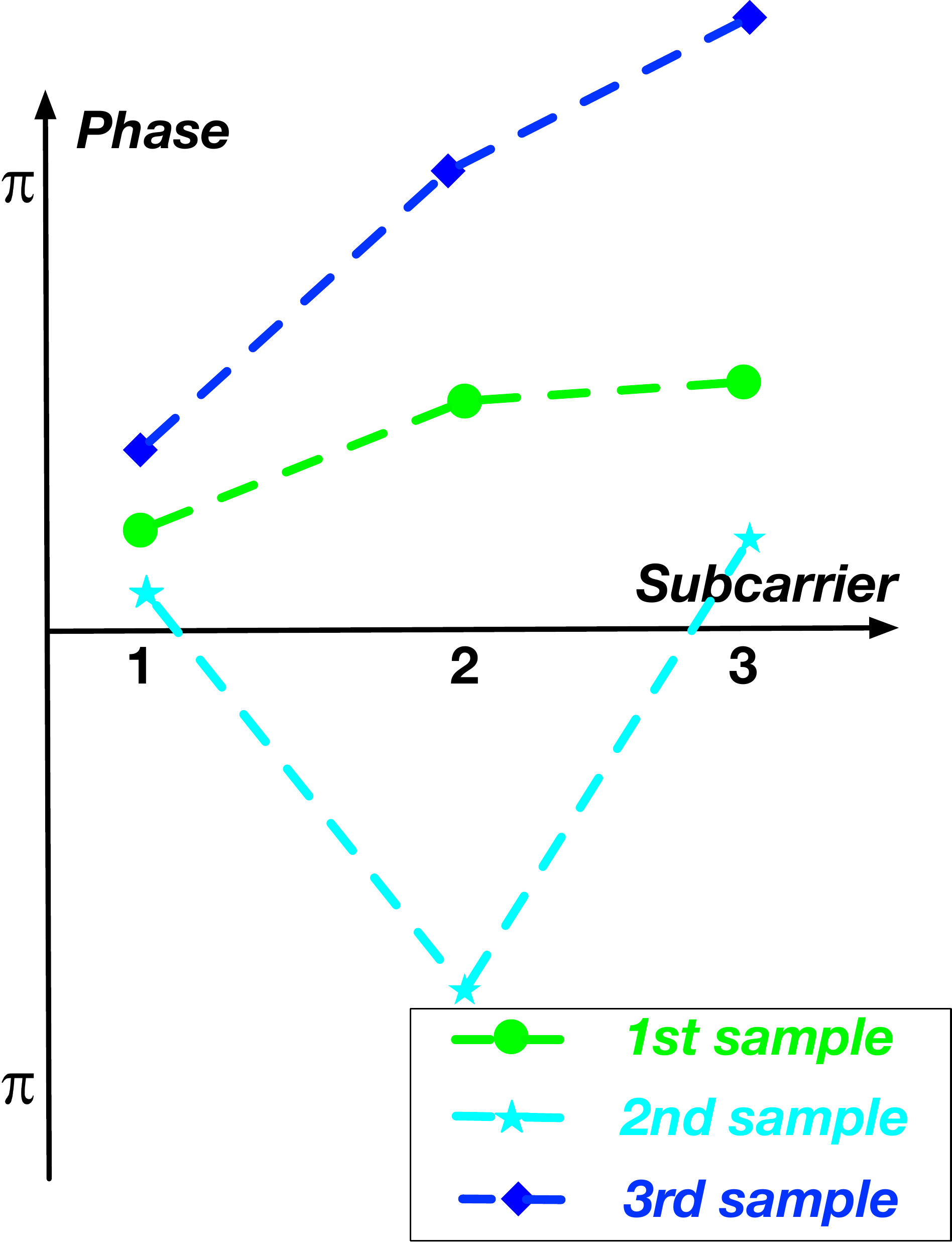}
        \label{fig:phase_unwrap}
    }
    \subfigure[Phase after sanitisation]{
         \includegraphics[scale = 0.25]{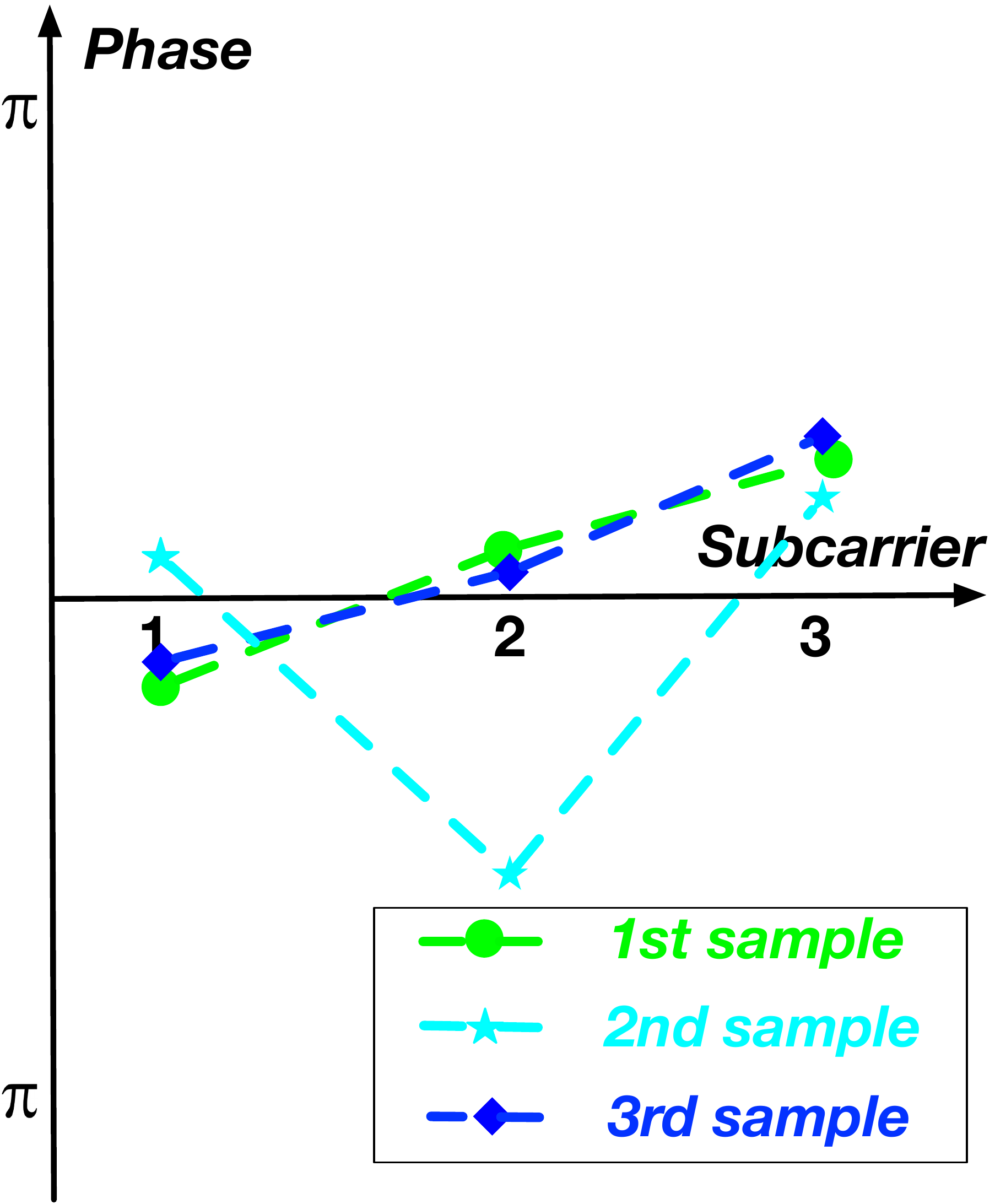}
        \label{fig:phase_santinised}
    }
    \caption{Examples of data processing - Part 3}
   \label{fig:3samples-p3}
\end{figure*}

\begin{figure*}[!ht]
    \centering
    \subfigure[Phase Shift]{
         \includegraphics[scale = 0.25]{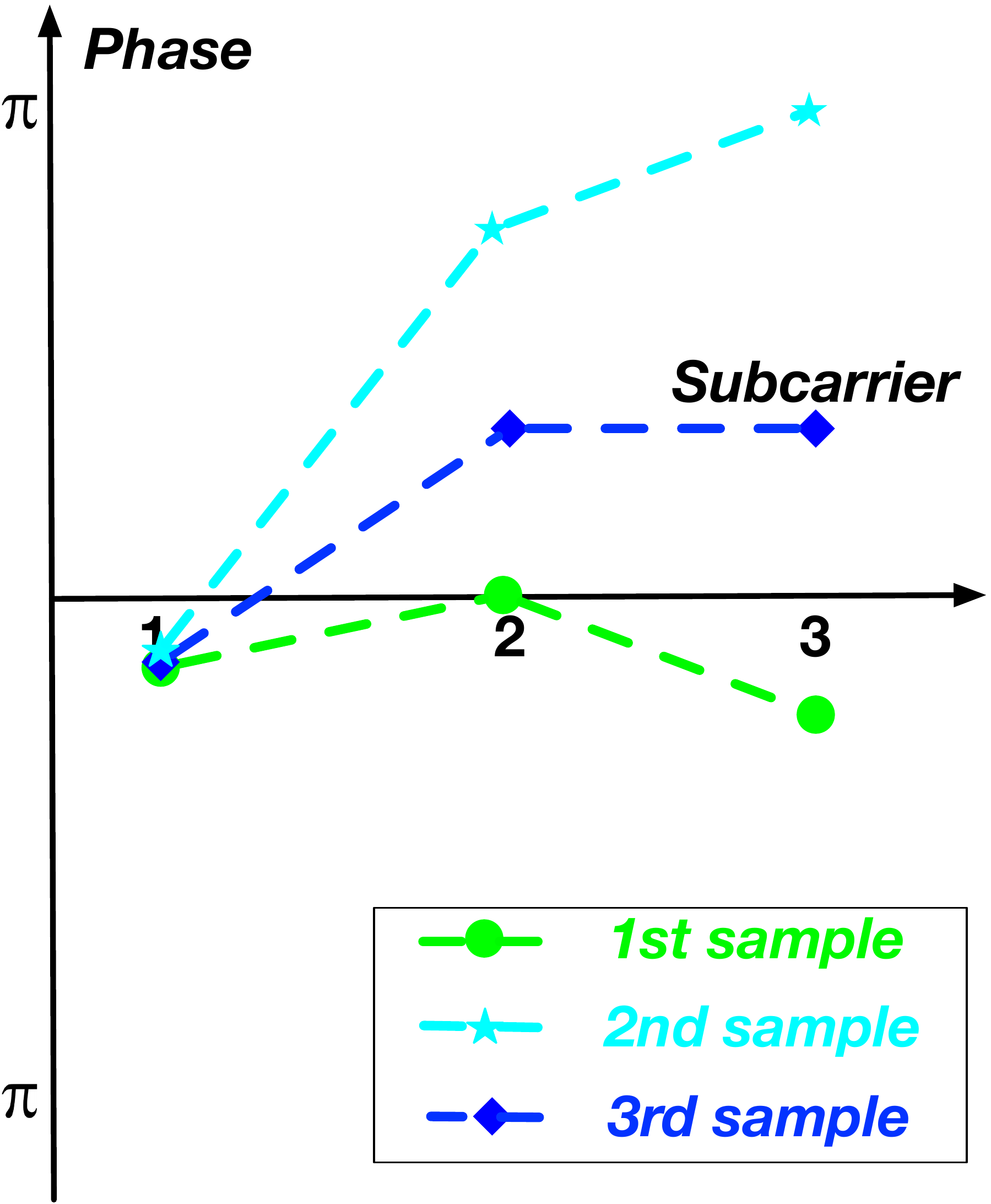}
        \label{fig:phase_shift}
    }
    \subfigure[The confusion of phase change]{
         \includegraphics[scale = 0.25]{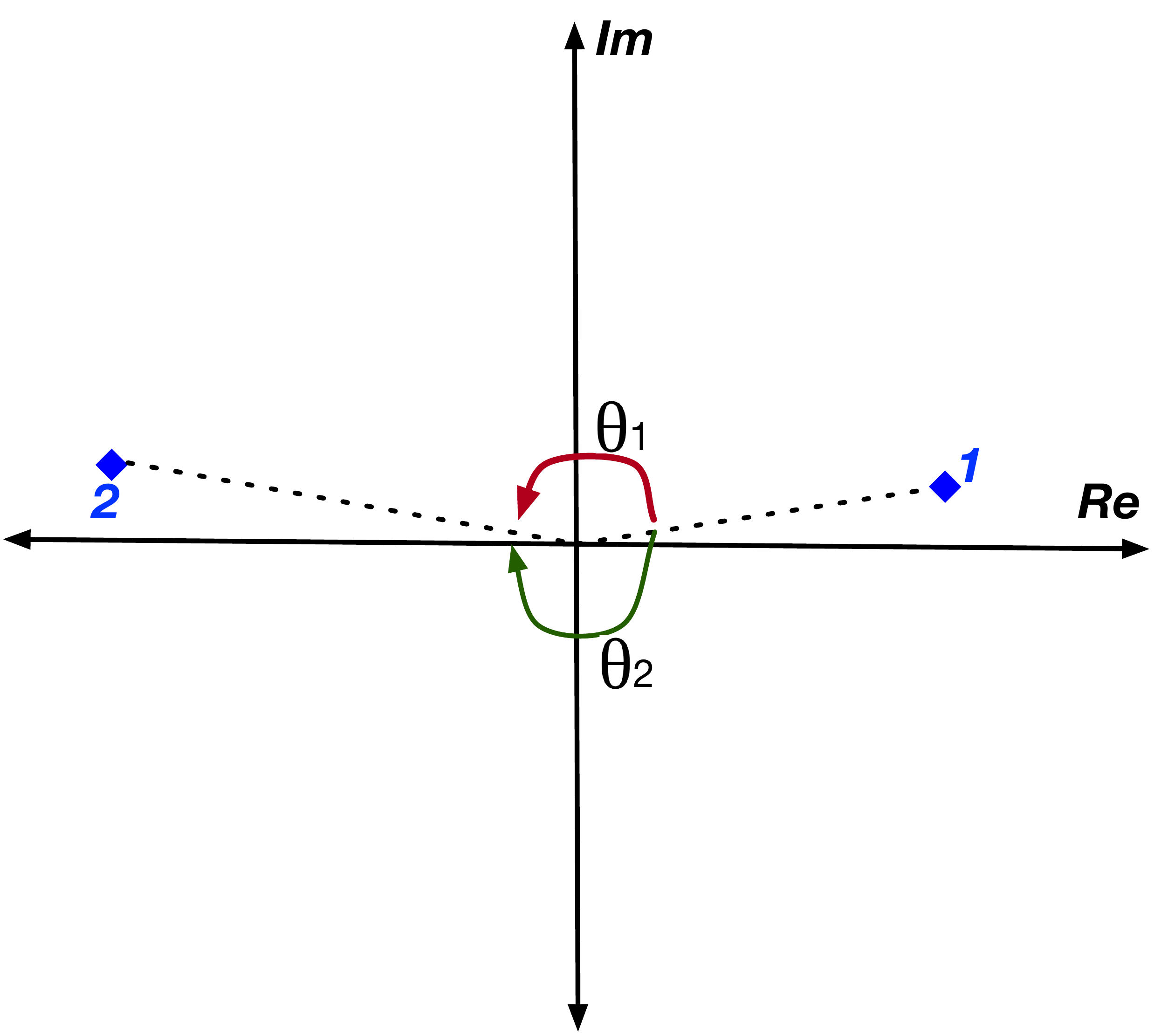}
        \label{fig:phase_confuse}
    }
    \caption{Examples of data processing - Part 4}
   \label{fig:3samples-p4}
\end{figure*}

\section{Background and Motivation} 
\label{sec:background}
The WiFi physical layer uses a multi-carrier modulation method called Orthogonal Frequency-Division Multiplexing (OFDM). At the 2.4GHz or 5.8GHz WiFi band, the bandwidth of one channel is 20 MHz, and each channel has 52 subcarriers \footnote{30 of the total 52 subcarriers in each channel at the 2.4GHz WiFi band can be reached when using the modified firmware of Intel 5300 WiFi card. }. 
The frequency response and phase shift of the centre frequency of each subcarrier are contained in CSI. 
The context variation significantly differentiates the multipath propagation of wireless signals and can be well reflected by CSI. 
The sensitivity of CSI to surrounding environments is broadly explored for context awareness by using supervised machine learning techniques.  

Specifically, to make a prediction, supervised machine learning techniques require an explicit trained model obtained at the training stage using labelled data. The good quality of training data is exceptionally beneficial for increasing prediction accuracy. Two main factors prominently affect the quality of training data, i.e. the distance between classes and within one class.
The high-quality training data further enlarge the distance between classes and narrow the distance within one class\cite{sun2010analysis,wei2019from}, which plays a vital role in improving recognition accuracy.

One CSI measurement is contained in a complex-valued vector $C = [C_1, C_2, ... C_n]$, where $n$ is the number of subcarriers. 
To better illustrate the steps of CSI signal processing, we select 3 neighbouring subcarriers among them as an example. 3 CSI samples are continuously collected in a very short period during a static gesture is performed in order to ensure no changes in the surrounding environment. The raw CSI measurements in the complex domain are shown in Figure~\ref{fig:complex_plane}.
Figure \ref{fig:amplitude} shows the amplitudes of these 3 CSI samples. 
Many radio-based applications conduct context prediction only using CSI normalised amplitude information \cite{wang2014eyes,wei2015radio}. 
Amplitude information is the absolute values of each CSI subcarrier, able to provide relatively good performance for many radio context awareness applications.
Normalisation is also applied because of the RSS variation as the example shown in Figure~\ref{fig:amplitude}.
Figure~\ref{fig:amplitude_normalised} demonstrates the constancy of normalised CSI amplitude information for a static context. 
However, the lack of phase information degrades recognition performance in many complicated scenarios, such as context awareness in radio frequency interference \cite{wei2019from} or with a large number of classes \cite{ma2018signfi}, where complementary phase information need to be explored.
\emph{Due to a random phase shift in OFDM from the unsynchronised time between the transmitter and receiver, the phase information, however, has not been directly used without explicitly processing in these applications.}
Sen et al. \cite{sen2012you} proposed a CSI data sanitisation method to calculate the phase information, which is massively used by phase-based or complex-valued CSI context awareness applications\cite{ma2018signfi,wei2018real,wang2015deepfi}. The sanitisation method firstly unwraps the raw phase and calculates the phase slope and offset to eliminate the phase shift. Complex-valued vectors can be constructed using the normalised amplitude and phase vectors. 
The step of unwrapping corrects aims to correcting radian phase angles to avoid the jump between consecutive subcarriers. In the unwrapping process, when the phase difference between neighbouring subcarriers is above the threshold (the default threshold is $\pi$), a compensation $2\pi$ is added/subtracted to the phases of its following subcarriers. 

%
%
%


CSI based context awareness commonly relies on supervised machine learning techniques. 
Different contexts derive multiple radio transmitting paths resulting in distinguishable CSI patterns, while the same context should be indicated by similar CSI patterns. 
However, improper signal processing may introduce unnecessary errors. 
Take unwrapping CSI phases as an example, the phase difference of many consecutive subcarriers is extremely close to the unwrapping threshold.
Unwrapping method is applied to the samples of which the phase difference of these consecutive subcarriers is above the threshold, while the rest samples stay unchanged. 
It may introduce notably inconstant situations for these consecutive CSI samples within one context because not all of those neighbouring subcarriers in samples are treated by unwrapping.  
Figure \ref{fig:phase_raw} are Figure \ref{fig:phase_santinised} are examples to further imply the unnecessary unreliability of CSI measurements caused by the unwrapping step. 
Phases for the three samples in Figure \ref{fig:complex_plane} without any processing are shown in Figure \ref{fig:phase_raw}. 
The phase difference between subcarriers 1 and 2 of sample 3 is above the threshold $\pi$.
Therefore, after applying the unwrapping approach, the phases of sample 3 are unwrapped, and the phases of samples 1 and 2 stay unchanged, as shown in Figure \ref{fig:phase_unwrap}. 
Please note the similar shapes of the raw samples 2 and 3 become completely different after unwrapping step.
Except for the unavoidable noise, another reason to result in this unwrapping issue is the changes in slope due to the asynchronous clock, which is demonstrated by another example shown in Figure \ref{fig:phase_shift}.
The phases of these 3 samples have the same shape with different orientations. Please note the phase difference between subcarriers 1 and 2. Its difference in sample 2, above the default threshold $\pi$,  is much greater than that in sample 1. 
As a result, the phase from sample 2 will be the only case of being unwrapped when processing their raw phases using the unwrapping method.
The unwrapping step produces unnecessary errors when applying supervised machine learning techniques on those samples because supervised machine learning resorts to the repeatable patterns from training data.  
The third reason for the instability at the unwrapping step is the different rotation directions in the complex domain as shown in Figure \ref{fig:phase_confuse}. 
$\theta_1$ and $\theta_2$ indicate phases for the same vector with anticlockwise and clockwise rotation in the complex domain, but $\theta_2$ is above the default threshold for unwrapping.
To summarise, unwrapping in the CSI sanitisation method makes the phase information feasible for context awareness but introduces unnecessary unreliability.  

To address these issues, we propose a deep learning based general framework which considers the additional informative phase information along with the amplitude information to improve CSI based context recognition performance and avoid the unnecessary unreliability by pruning any initial data pre-processing stage.

%

%% file: Tex/method.tex

\begin{figure*}[!ht]
    \centering
    \subfigure[Traditional DNN model]{
         \includegraphics[scale = 0.25]{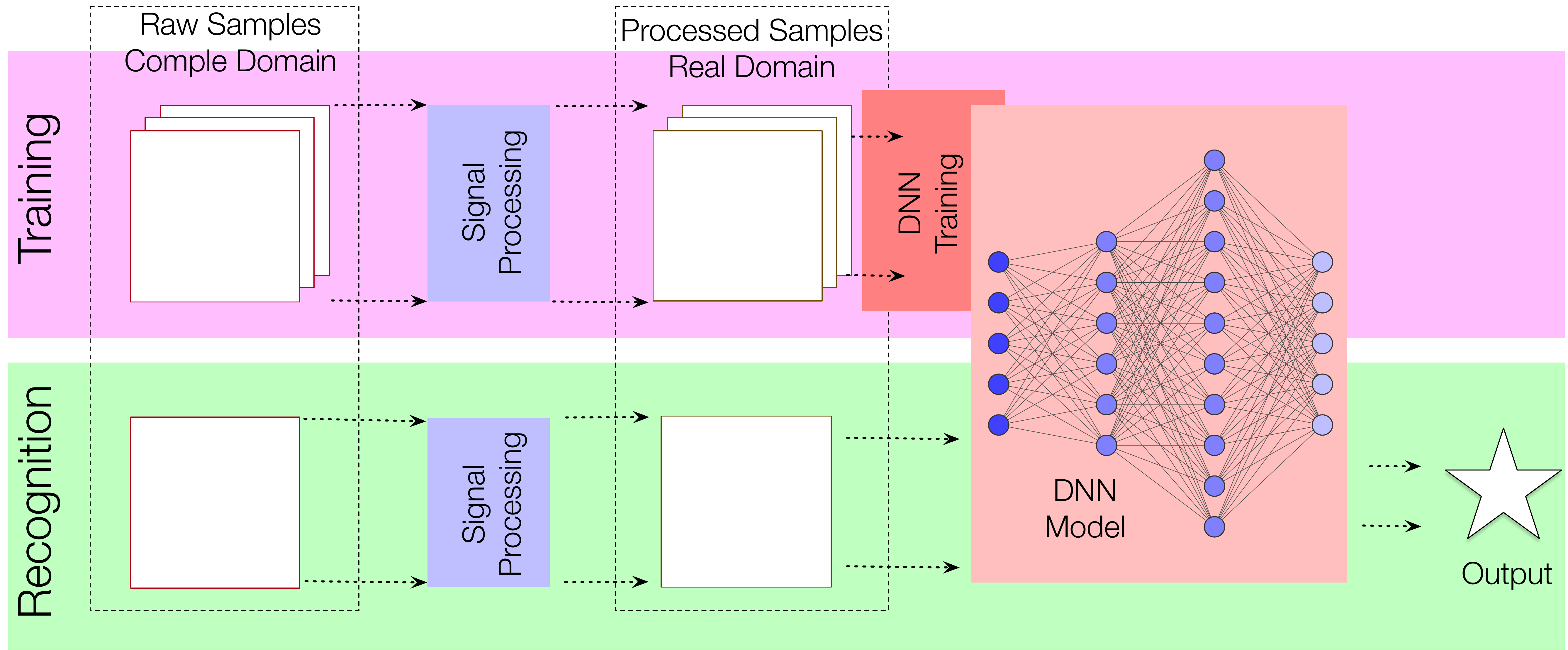}
       \label{fig:trainining_test_traditional}
    }
    
    \subfigure[Our proposed Model]{
         \includegraphics[scale = 0.25]{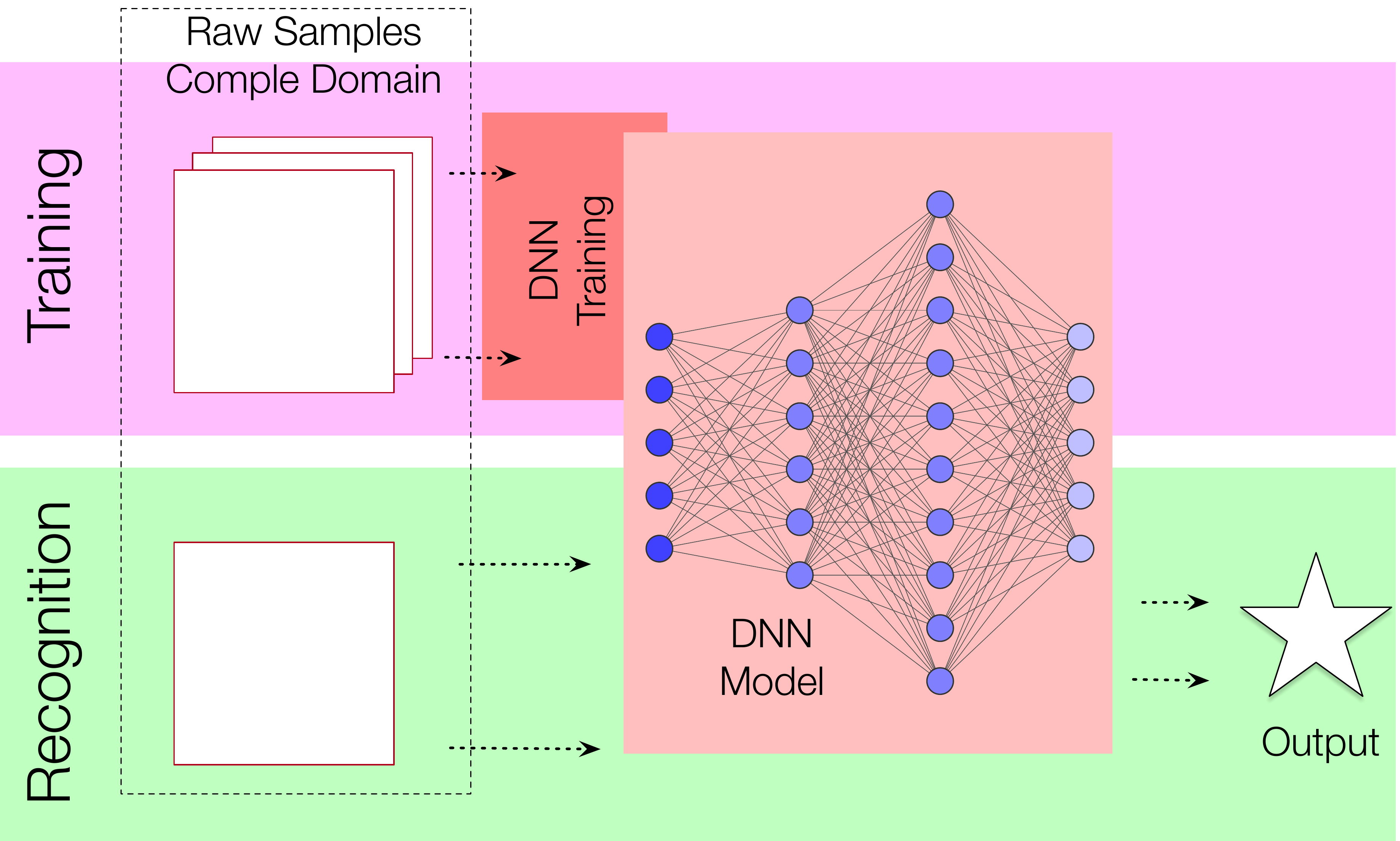}
        \label{fig:trainining_test_proposed}
    }
    \caption{DNN models}
   \label{fig:dnn_model}
\end{figure*}


\begin{figure*}
   \centering
    \includegraphics[scale = .2]
    {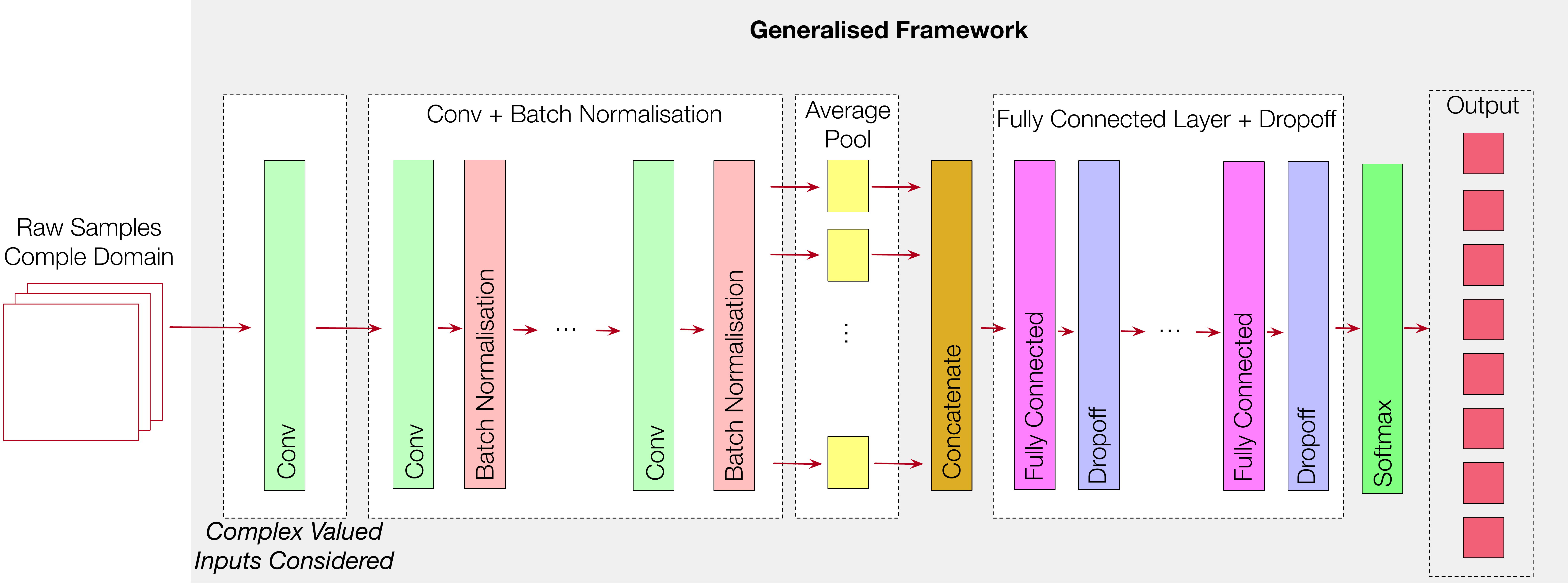}
    \caption{The general framework for radio-based device-free context awareness}\label{fig:CNN_diagram}
\end{figure*}

\begin{figure*}
   \centering
    \includegraphics[scale = .8]
    {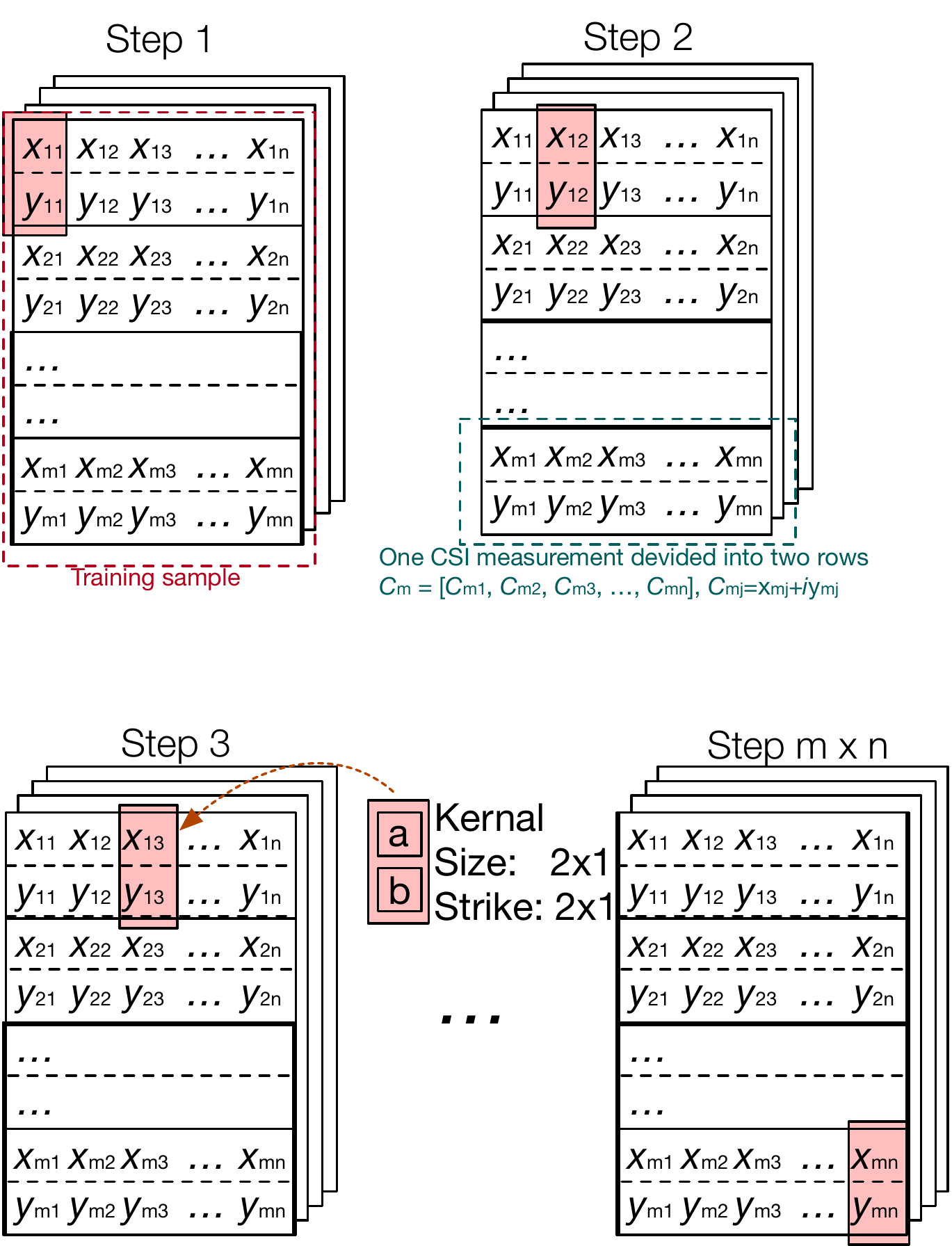}
    \caption{first layer kernal}\label{fig:first_layer_kernal}
\end{figure*}
\section{General Framework: CSI based Context Awareness Using Deep Learning}
\label{sec:methods}
The sanitisation process in \cite{sen2012you} tries to regularise the phase and constructs complex valued CSI measurements for further classification and clustering. 
Traditionally, computer vision recognition algorithms also use similar strategies. Data representation is explicitly designed for specific classification methods to improve the recognition performance, such as features in the face recognition field: Eigenface\cite{zhang1997face}, Fisherface\cite{kwak2005face} and Laplacianface\cite{he2005face}.  
Recently, researchers use deep learning techniques to train DNN models using raw images without any pre-processing instead.  
The application of deep learning has significantly enhanced performance in many domains, such as face recognition\cite{parkhi2015deep}, object recognition\cite{lecun2015deep}, etc. It can preserve all the features and avoid unnecessary information loss by directly processing raw images with deep learning techniques. 
They also propose a framework which can be generalised to other datasets \cite{simonyan2014very}.
%

Motivated by this, we propose a general framework and investigate the application of raw CSI measurements for radio-based context awareness applications without any signal processing. 
The advantages are two-fold. Firstly, the unmodified CSI measurements preserve all the original and relevant information, and the DNN helps select the desirable features for context awareness instead of designing feature representation. 
Secondly, the framework can be applied to all the CSI based context awareness applications because our proposed algorithm does not design features for specific application scenarios. 
Several research works have already applied deep learning on radio-based context awareness, but they have an additional step to process raw CSI measurements for training a DNN model. Furthermore, they have not investigated the feasibility of adapting their models to other applications or datasets. Figure \ref{fig:trainining_test_traditional} shows the structures of those traditional applications. 
Firstly, they apply a data pre-processing method to prepare the raw complex-valued CSI samples in the complex domain for further classification. There are two common signal processing methods for CSI data: (1) abstract the amplitude of complex-value CSI measurements by calculating absolute values of that in each subcarrier; or (2) calculate the phase using the sanitisation method introduced in \cite{sen2012you}, and use phase information to construct sanitised complex vectors along with the amplitude information or as complimentary of the amplitude information. The processed measurements are taken into consideration for training a DNN model.

Different from the traditional methods, we propose a DNN model based general framework by using the raw CSI measurements in the complex domain directly. 
To achieve this, our proposed framework uses the structure shown in Figure \ref{fig:trainining_test_proposed}. Instead of initially processing raw CSI measurements, we explore them directly for DNN model training and inference process. 
As illustrated in Figure \ref{fig:CNN_diagram}, our proposed general framework has the following important components in the neural network architecture: (1) convolutional layers, (2) batch normalisation layers, (3) average pooling layers (4) dropout layer (5) fully connected layer and (6) softmax layer. Here are more details of our proposed architecture.

\noindent \textbf{\emph{Raw} CSI measurements as Input:} For CSI based context awareness, the shape of each instance is 3 dimensions, i.e. $2m \times n \times c$. $m$ is the number of CSI measurements in one CSI instance. We use real and imaginary parts together for training our deep model, so the size of this dimension is $2m$. $n$ and $c$ are the numbers of subcarriers and antenna pairs, respectively. Figure~\ref{fig:first_layer_kernal} shows the example of one instance.  

\noindent \textbf{Convolutional Layer and Batch Normalisation Layers:} 
We use multiple convolutional layers in our framework. The first layer considers the complex-valued inputs. Figure~\ref{fig:first_layer_kernal} shows each step when using Kernal whose size and strike are both $2 \times 1$. This layer takes both the complex and real parts of the CSI measurements into accounts. The strike $2 \times 1$ ensures this layer only considers the relevant subcarrier of each sample at one time. 
The followed convolutional layers use different size of kernels to explore the diversity of the features.  
Batch normalisation layers are also used along with convolutional layers. The inputs of DNN are raw CSI instances without any initial data processing, so the batch normalisation layers can scale the activations in order to accelerate the training process and improve the performance. The batch normalisation layers are also important in managing extremely diverse radio signal strength.

\noindent \textbf{Average Pooling Layer and Concatenate Layer:} 
After extracting features from convolutional layers, we take the neighbouring subcarriers and samples into consideration and use average pooling layers. This layer plays an essential role in recognition of various degrees of movements. As shown in Table~\ref{tab:con_dnn}, in this layer, we used concurrent multiple pool sizes in order to consider different ranges of subcarriers generated from different contexts. 
A concatenate layer is appended to combine features from the outputs of these average pooling layers.

\noindent \textbf{Dropoff Layer:}
Overfitting is a common issue in the machine learning domain when the evaluation performance decreases with the increase of the training performance. The dropoff layer is a simple but effective technique to address the overfitting issue in deep learning based methods, which drop a certain percentage of the outputs from the previous layer. Dropoff layers are also employed in our framework to avoid the overfitting in the training stage. 

\noindent \textbf{Fully Connected Layer:} 
Fully connected layers, along with ReLU non-linearity activation, are introduced to assemble the features from the above layers and determine patterns for further decision making. 

\noindent \textbf{Softmax Layer:} The final softmax layer is constructed to recognise context based on the hidden state from fully connected layers.

This paper aims to propose a general framework for radio-based device-free context awareness, and the proper configuration for each particular application needs to be considered. The following Section \ref{sec:experiments} will evaluate our proposed general frameworks with specific DNN structures in two application scenarios. 

%% file: Tex/experiment.tex
\section{Performance Evaluation: The applications of the Proposed General Framework} \label{sec:experiments}

\begin{table}[]
\caption[Table caption text]{Radio Based Gesture Recognition Dataset Overview}\label{tab:dataset}
\begin{tabular}{lllll}
\hline
Dataset ID & Environment & \begin{tabular}[c]{@{}l@{}}Number of \\ Gestures\end{tabular} & \begin{tabular}[c]{@{}l@{}}Number of \\ Users\end{tabular} & \begin{tabular}[c]{@{}l@{}}Number of \\ Instances\end{tabular} \\ \hline 
D1         & Home        & 276                                                           & 1                                                          & 2,760                                                          \\ \hline
D2         & Lab         & 276                                                           & 1                                                          & 5,520                                                          \\ \hline
D3         & Lab         & 276                                                           & 1                                                          & 5,520                                                          \\ \hline
D4         & Lab         & 150                                                           & 5                                                          & 7,500                                                         \\ \hline
\end{tabular}
\end{table}

In this section, we validate our proposed deep neural network based framework based on two applications to show its generalisation and performance: (1) \emph{SignFi}: the public datasets from SignFi \cite{ma2018signfi} for gesture recognition \footnote{The datasets can be downloaded from https://yongsen.github.io/SignFi/} and (2) \emph{Activity}: the datasets for radio based activity recognition with radio frequency interference in \cite{wei2019from} \footnote{The preprint version of this paper can be found in https://arxiv.org/abs/1804.09588}. 
In this section, we show the DNN structures for each application based on the general framework. 
We also compare our proposed model with existing methods and show the effect of the convolutional layer, the batch normalisation layer, the average pooling layer and fully connected layer.

\subsection{The DNN Structure and Performance in SignFi Datasets}

\subsubsection{Dataset Description}

There are four public datasets in SignFi obtained from experimental measurements for radio-based gesture recognition in a laboratory and a home scenario. The datasets contain raw CSI measurements with regards to different room sizes, distances between the transmitter and receiver, and orientations of their antennas. 
In these datasets, the sampling rate of CSI measurements is 200Hz, and the gesture duration in one instance is between 0.5 seconds and 2 seconds. 

The detailed information on the four datasets of SignFi is summarised in Table \ref{tab:dataset}. 
In D1, 276 gestures are performed by one user in the home scenario, and 2,760 instances are collected.
The CSI instances from D2 and D3 are simultaneously collected from the receiver and transmitter, respectively. The receiver and transmitter are deployed in the lab, where 5,520 CSI instances are collected. 
The CSI instances in D4 are also collected in the lab scenario, where 7,500 instances of gestures are performed by 5 users. Please note, due to multiple users in D4, each user's gestures are recognised by using his or her own CSI instances so that the DNN can be trained more accurately.

\begin{table}[]
\caption[Table caption text]{Configuration of DNN used in SignFi datasets}\label{tab:con_dnn}
\centering
\begin{tabular}{lcc}
\hline
                & Kernel Size         & Stride       \\ \hline
Conv\_1         & 2 x 1               & 2 x 1        \\
Conv\_2         & 3 x 3               & 1 x 1        \\
Conv\_3         & 5 x 5               & 1 x 1        \\
Conv\_4         & 10 x 10             & 1 x 1        \\ \hline \hline
                & \multicolumn{2}{c}{Pool Size}      \\ \hline
AP\_1           & \multicolumn{2}{c}{3 x 3}          \\
AP\_2           & \multicolumn{2}{c}{5 x 5}          \\
AP\_3           & \multicolumn{2}{c}{10 x 3}         \\
AP\_4           & \multicolumn{2}{c}{20 x 3}         \\
AP\_5           & \multicolumn{2}{c}{40 x 3}         \\ \hline \hline
                & \multicolumn{2}{c}{Dropout Rate}   \\ \hline
Dropout         & \multicolumn{2}{c}{0.8}            \\ \hline \hline
                & \multicolumn{2}{c}{Number of Unit} \\ \hline
Fully connected & \multicolumn{2}{c}{1000}           \\ \hline
\end{tabular}
\end{table}
\subsubsection{Architecture of DNN model}
To solve this problem, we use the general framework and the configuration of the model shown in Table \ref{tab:con_dnn} to achieve the best performance. Specifically, we use 4 convolutional neural network layers each with batch normalisation layer, and 5 average pooling layers, and 1 fully connected layer with dropout layer. 


In this section, we evaluate our proposed method 5-fold cross-validation with all the four datasets, as the same validation method is used in SignFi \cite{ma2018signfi} as well.
The datasets used by SignFi contains much more classes than other radio-based context awareness research works, which can be fully exploited to evaluate our proposed general framework. 
In \cite{ma2018signfi}, the authors proposed the deep learning model along with data pre-processing for radio-based gesture recognition. 
In this paper, we investigate the feasibility of using raw CSI without any data pre-processing instead.

When discussing the effect of the convolutional layers, the batch normalisation layers and average pooling layer, we only change the configuration of that layer category and the rest layers use the default model as shown in Table \ref{tab:con_dnn}.

\subsubsection{Comparing with SignFi}

\begin{figure}[!ht]
    \centering
    \subfigure[D1]{
         \includegraphics[scale = 0.33]{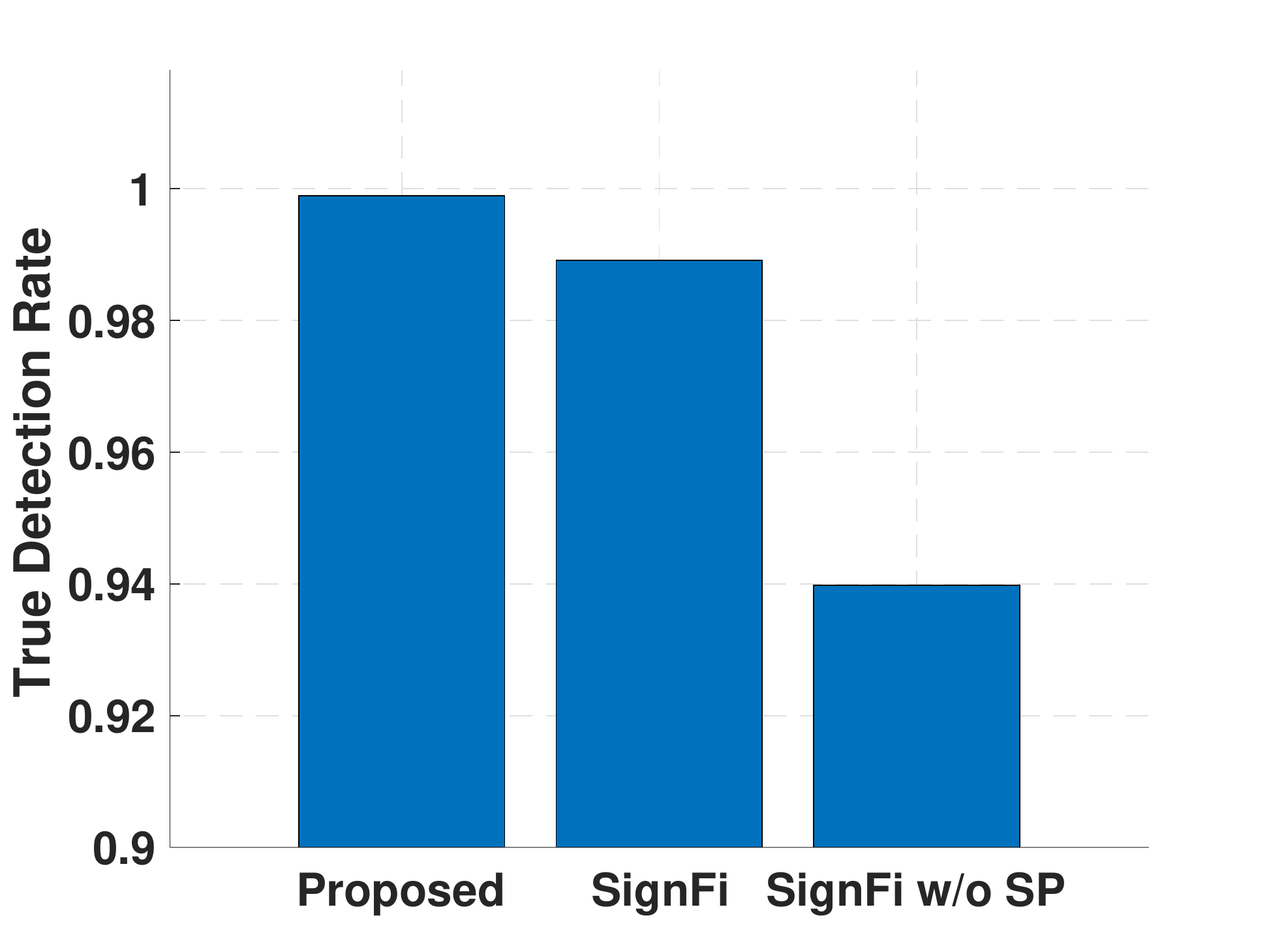}
        \label{fig:performance_D1}
    }
    \subfigure[D2]{
         \includegraphics[scale = 0.33]{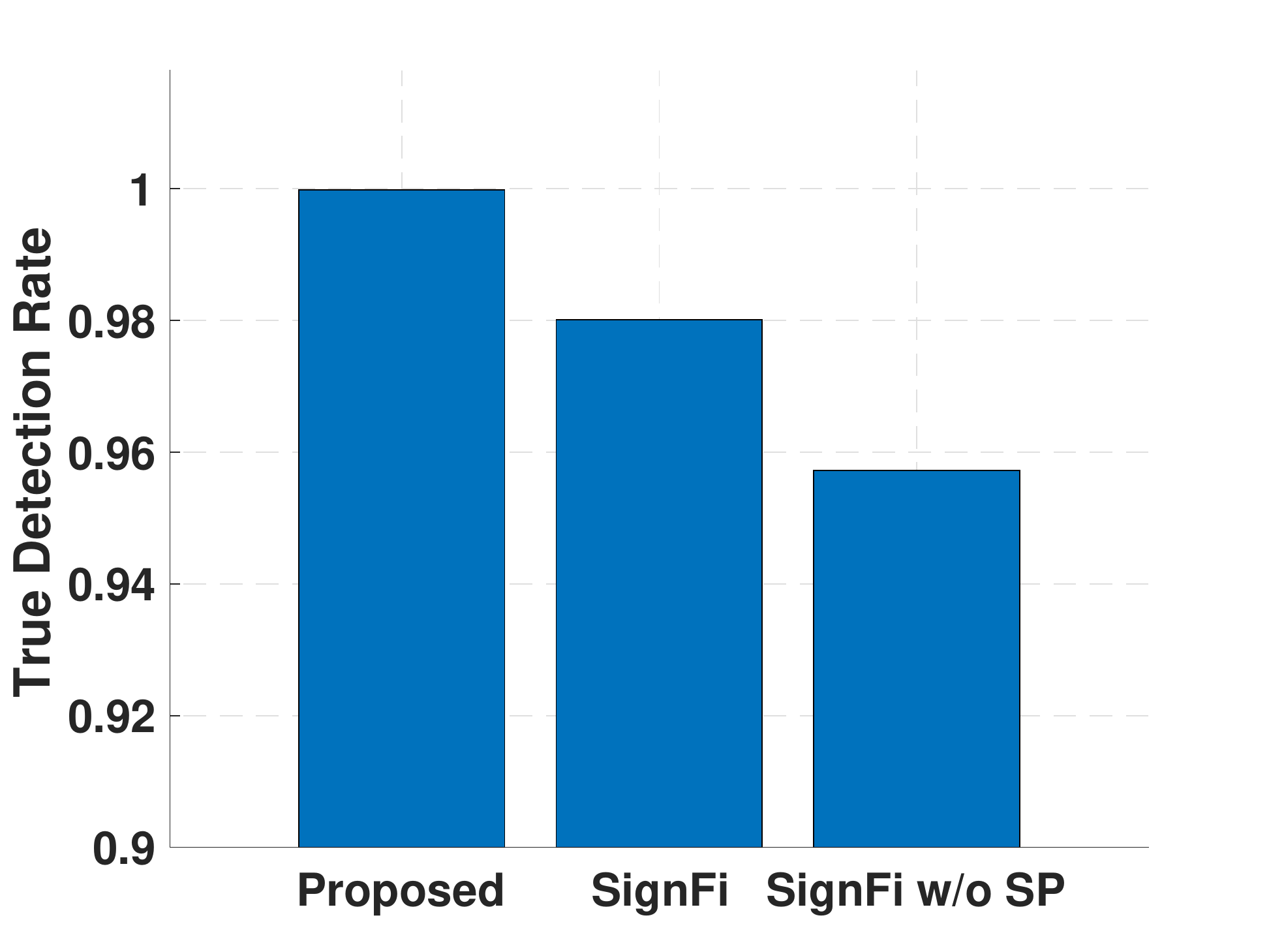}
        \label{fig:performance_D2}
    }
    
    \subfigure[D3]{
         \includegraphics[scale = 0.33]{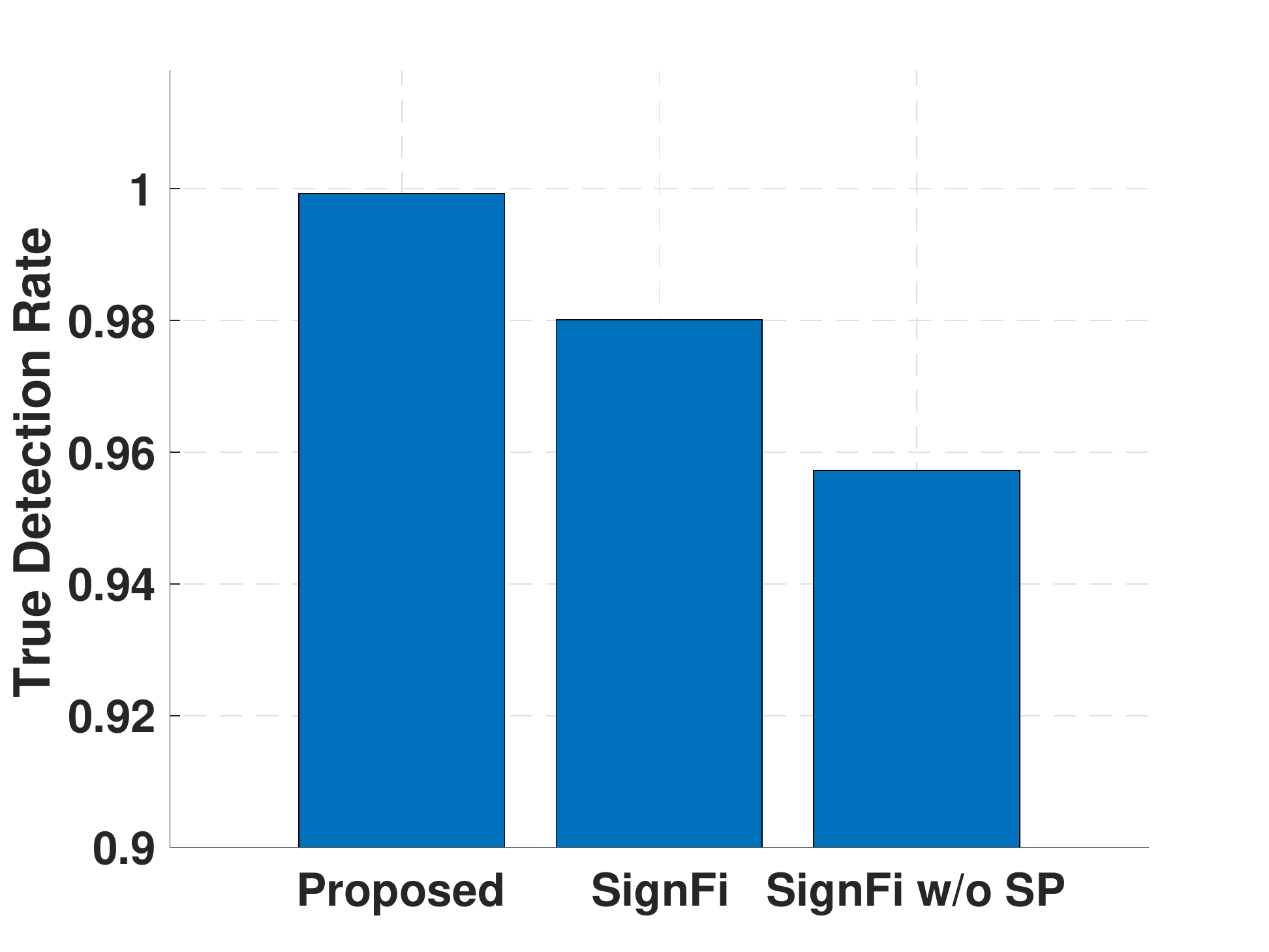}
        \label{fig:performance_D3}
    }
    \subfigure[D4]{
         \includegraphics[scale = 0.33]{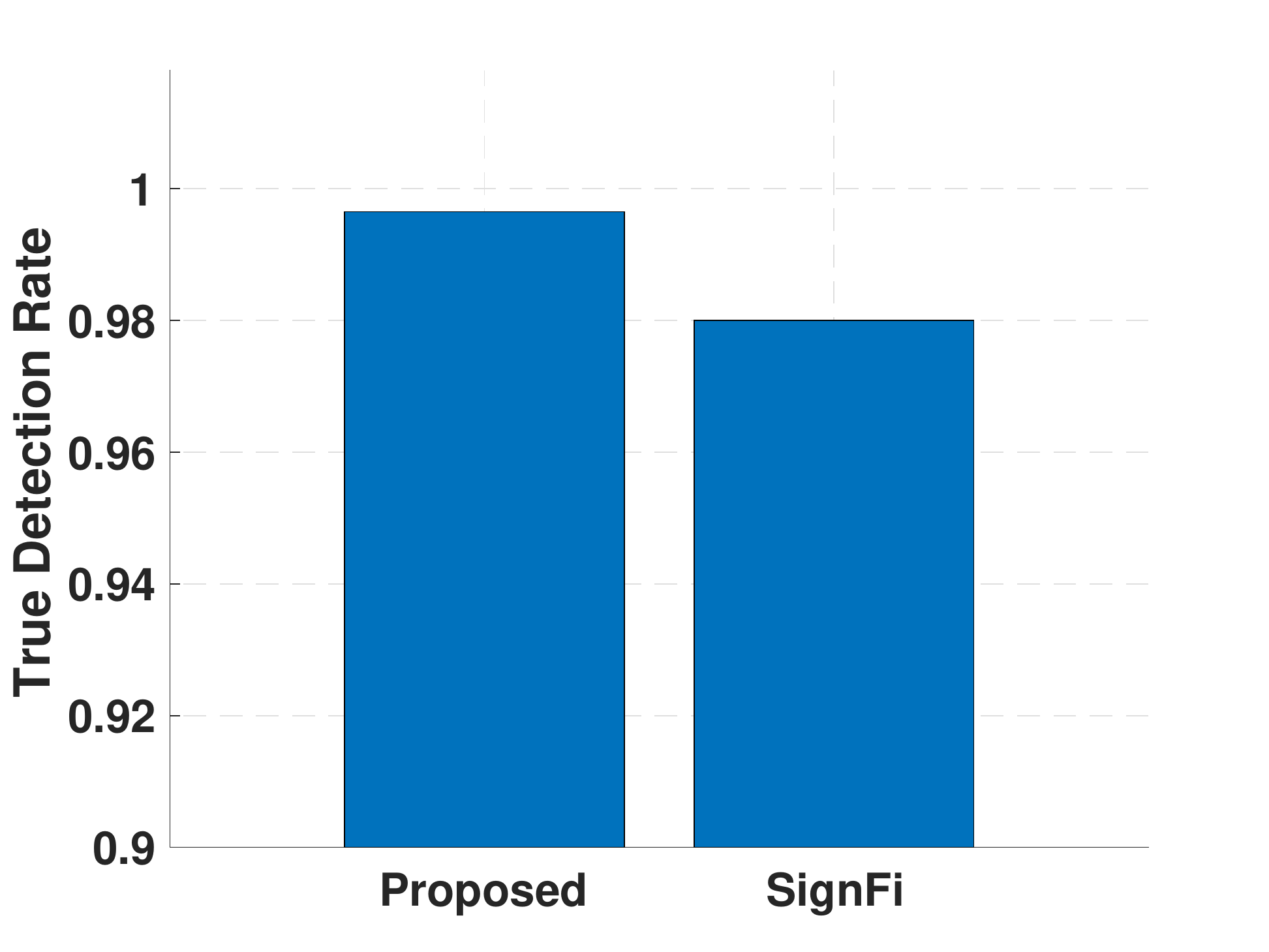}
        \label{fig:performance_D4}
    }
    \caption{The performance of true detection rate with regards to the four datasets.}
   \label{fig:performance}
\end{figure}

We compare our proposed DNN model with SignFi\cite{ma2018signfi} in terms of the true detection rate, which is defined as a ratio of the number of correctly recognised user's gestures and the total number of data samples. 
Figure \ref{fig:performance} shows the performance of our proposed method, and our DNN model achieves 99.89\%, 99.98\%, 99.93\% true detection rates in the home environment (D1) and the lab environment (D2 and D3), respectively.  
SignFi achieves 98.91\% and 98.01\% true detection rates in the home environment (D1) and the lab environment (D2 and D3 together), respectively. Our proposed deep learning model outperforms the existing model without any data pre-processing.
SignFi also discussed the performance of its deep learning model without pre-processing in the home environment (D1) and the lab environment (D2 and D3 together).
Without data pre-processing, the true detection rates drop to 93.98\% and 95.72\% using their method, while our proposed method can achieve nearly 100\% true detection rate. 
Since our proposed method does not require any additional data preprocessing, it can be readily extended to many CSI based applications. This will be further confirmed in Section \ref{subsec:exp_activity} that uses our proposed general framework for another application: radio-based device-free activity recognition in radio frequency interfered environment.


\subsubsection{Effect of the Convolutional Layer}
\begin{figure}[!ht]
    \centering
    \subfigure[D1]{
         \includegraphics[scale = 0.33]{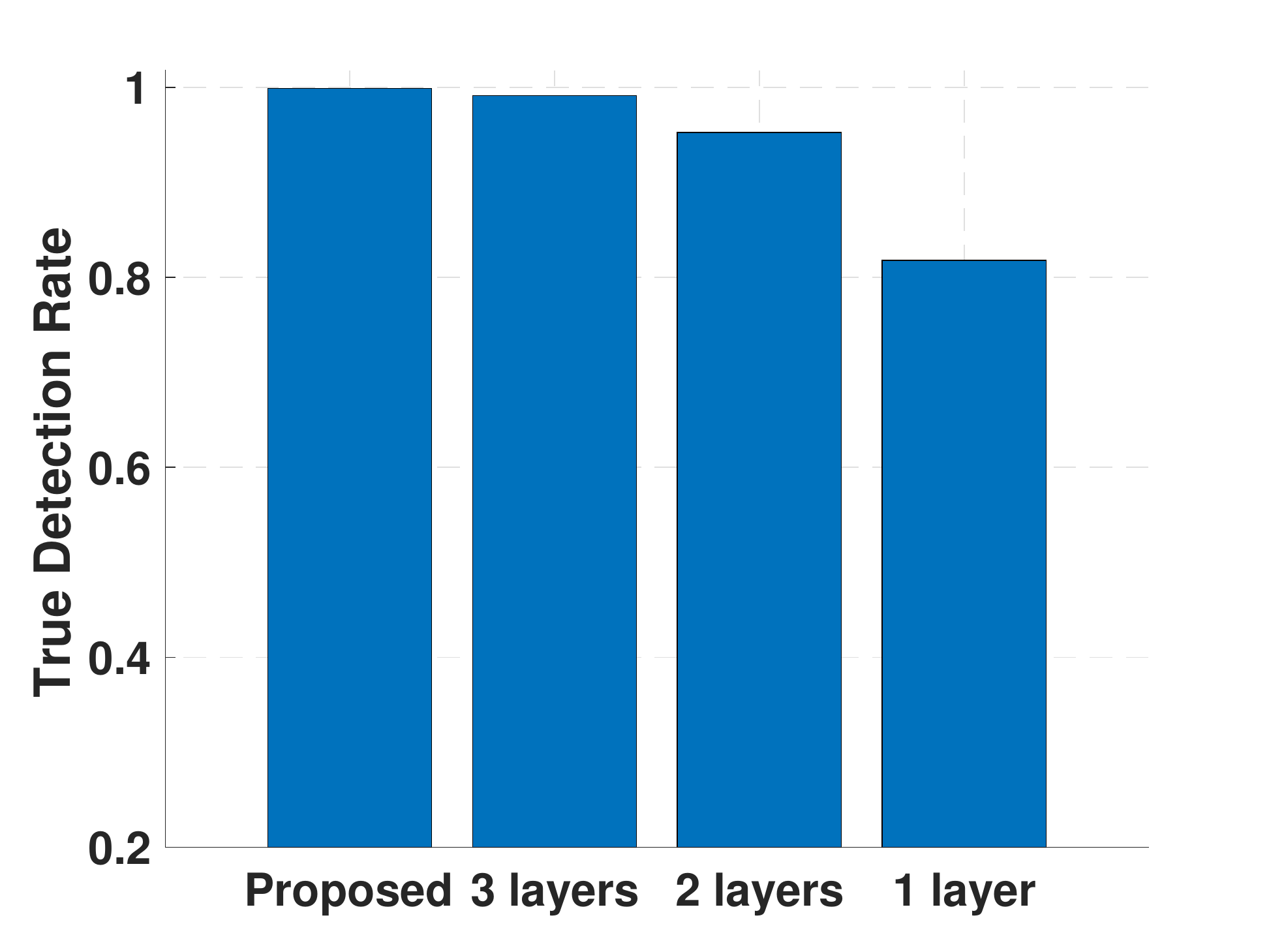}
        \label{fig:conv_D1}
    }
    \subfigure[D2]{
         \includegraphics[scale = 0.33]{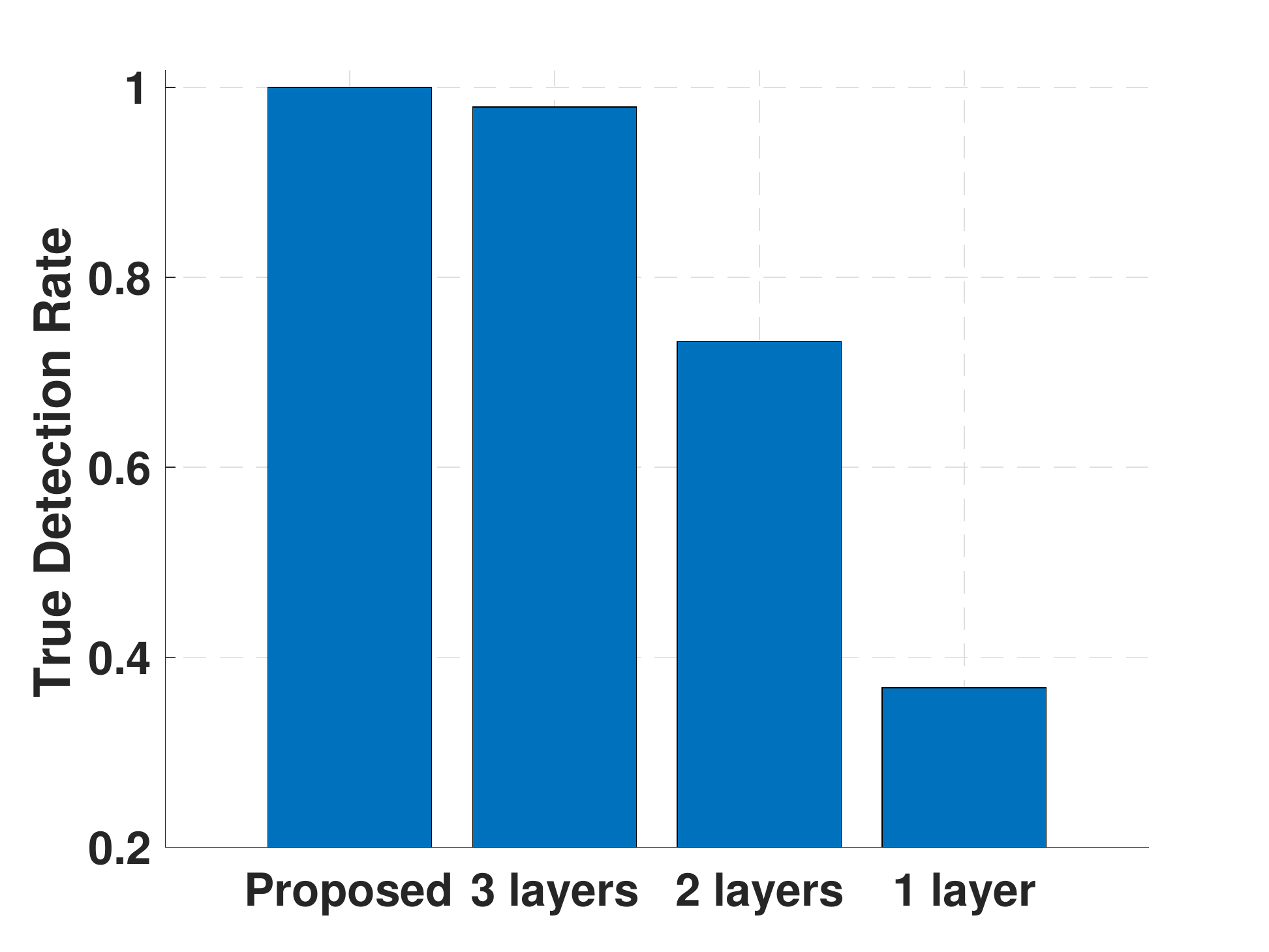}
        \label{fig:conv_D2}
    }
    
    \subfigure[D3]{
         \includegraphics[scale = 0.33]{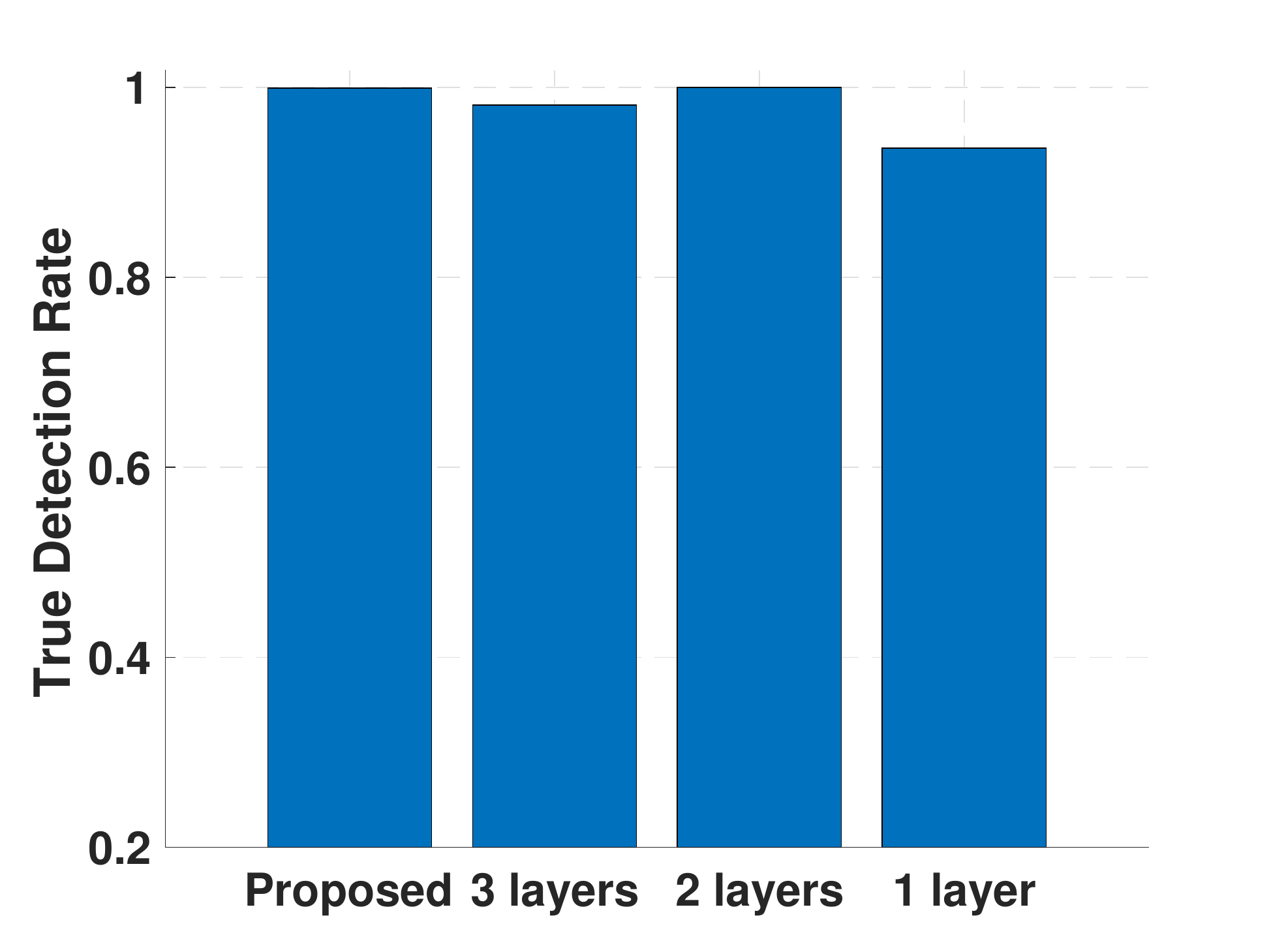}
        \label{fig:conv_D3}
    }
    \subfigure[D4]{
         \includegraphics[scale = 0.33]{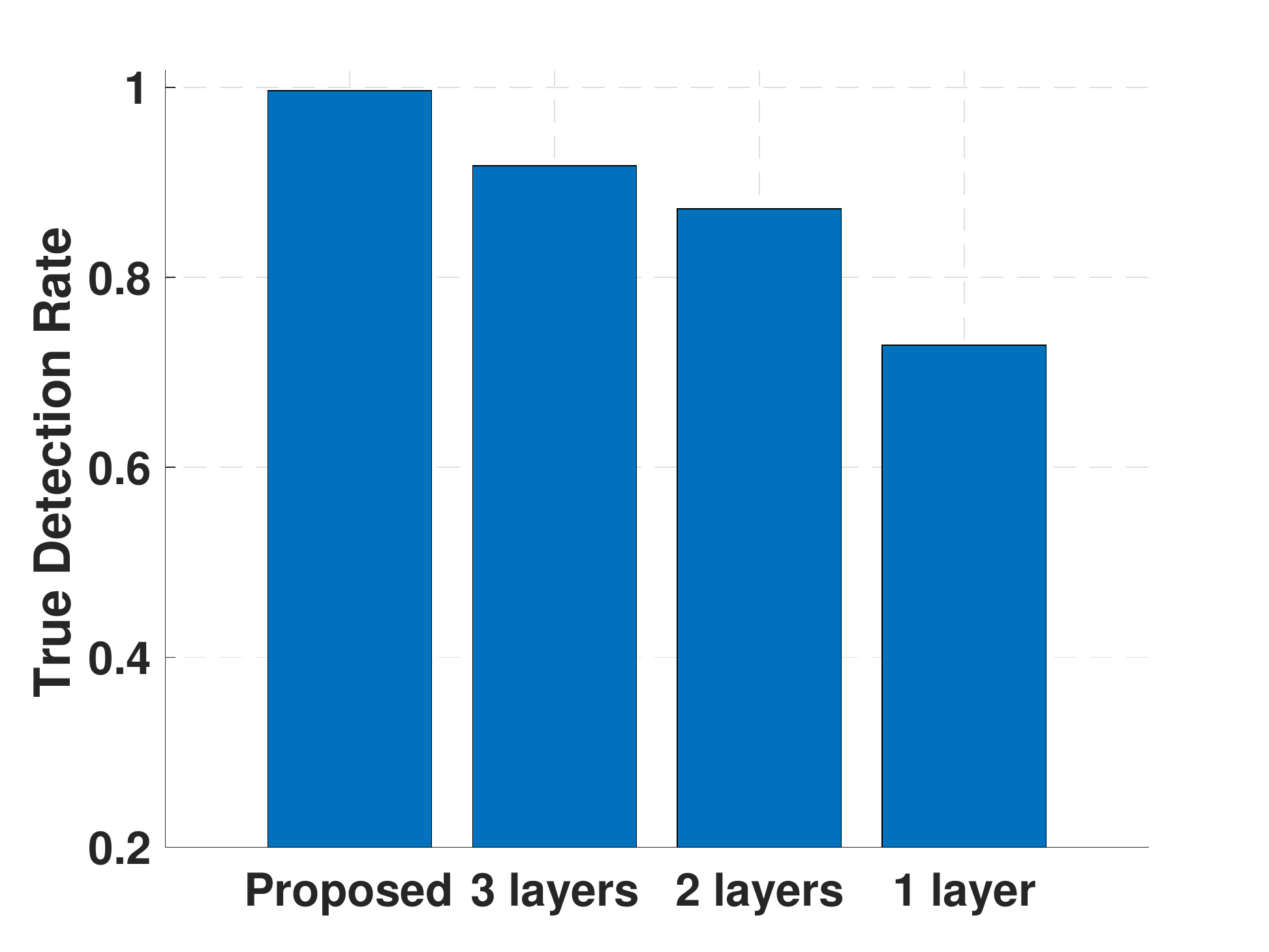}
        \label{fig:conv_D4}
    }
    \caption{The true detection rate of the proposed deep learning model with different number of Convolutional Layers}
   \label{fig:conv}
\end{figure}
Next, we investigate the effect of the number of convolutional layers. Figure \ref{fig:conv} shows the true detection rates of our proposed DNN model with regards to the number of convolutional layers, where the proposed model applies 4 convolutional layers.  
With using less convolutional layers, the performance drops dramatically except dataset D3. In D3, the true detection rate only drops from 99.93\% to 93.59\%. 
When using one less layer than the proposed model, the performance stays similar except the true detection rate drops 7.89\% from 99.65\% to 91.79\% in D4. 
For D2 and D4, the true detection rates are only 73.24\% and 87.23\% with 2 convolutional layers, and it furthers drops to 36.81\% and 72.88\% when only using 1 convolutional layer. 
These results confirm the increasing number of convolutional layers can significantly improve the accuracy, and our proposed model with 4 convolutional layers can achieve a sufficiently good and stable performance. 


\subsubsection{Effect of The Batch Normalisation Layer} \label{sec:exp_norm}
\begin{figure}[!ht]
    \centering
    \subfigure[D1]{
         \includegraphics[scale = 0.33]{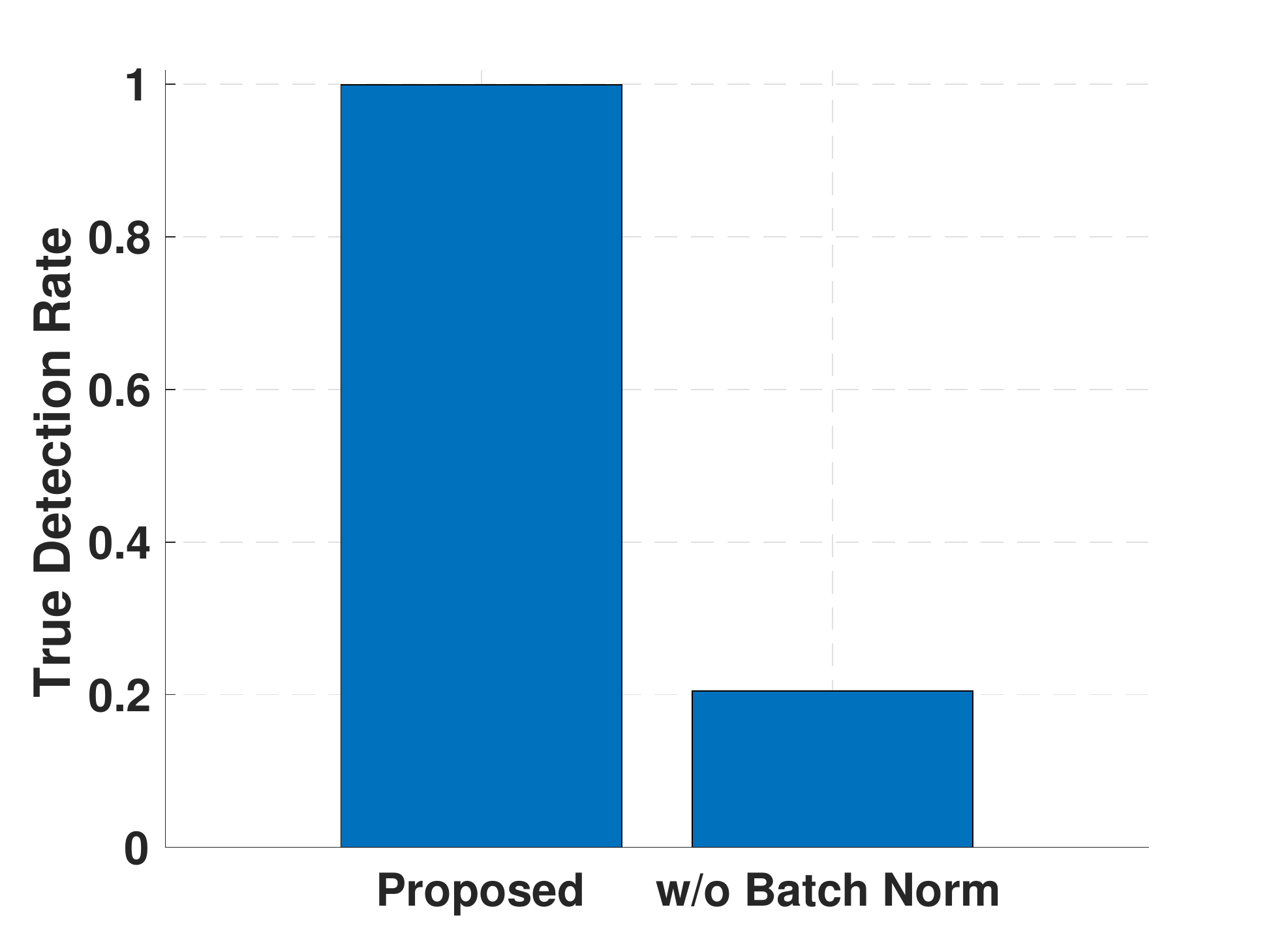}
        \label{fig:batchnorm_D1}
    }
    \subfigure[D2]{
         \includegraphics[scale = 0.33]{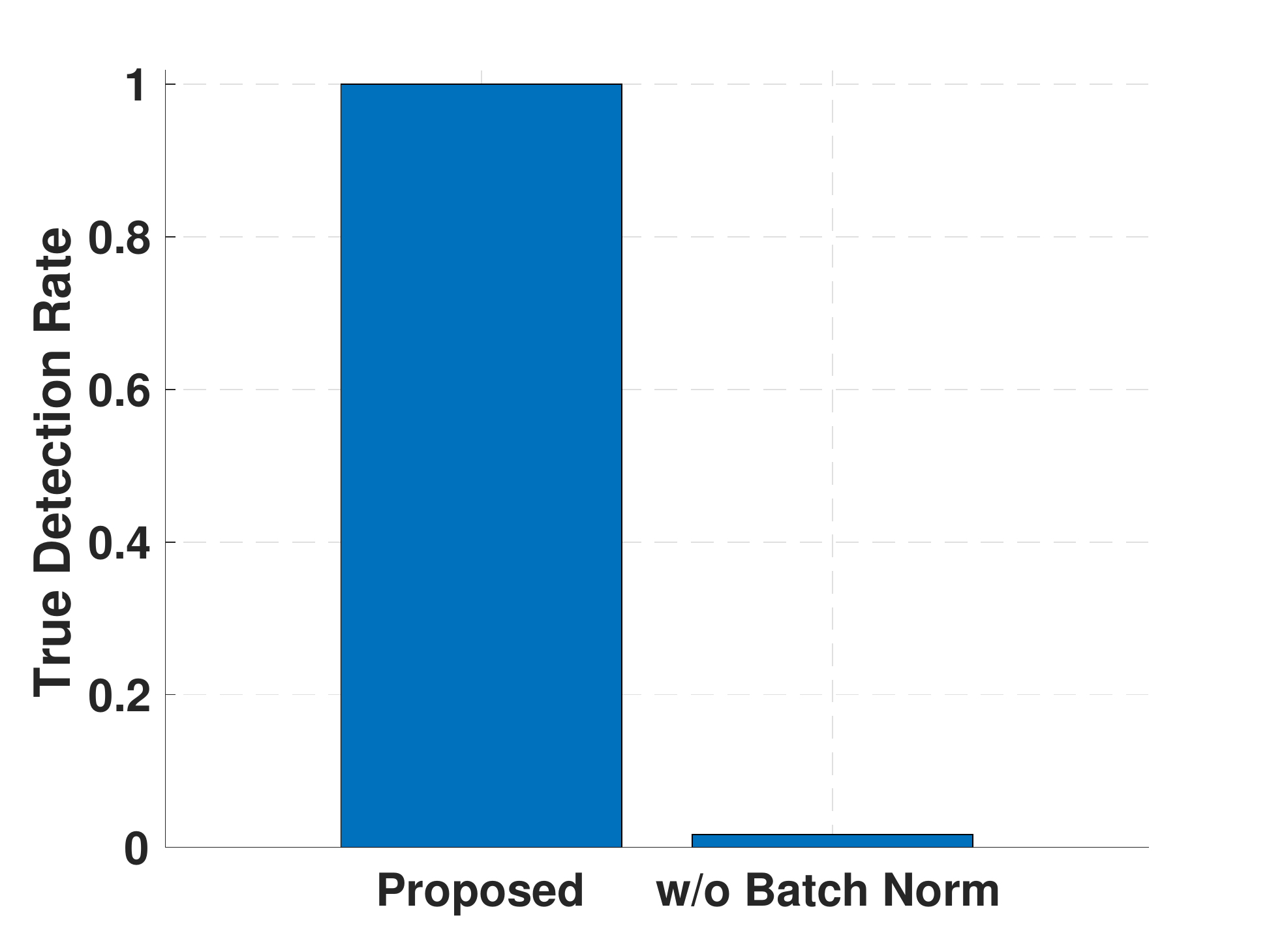}
        \label{fig:batchnorm_D2}
    }
    
    \subfigure[D3]{
         \includegraphics[scale = 0.33]{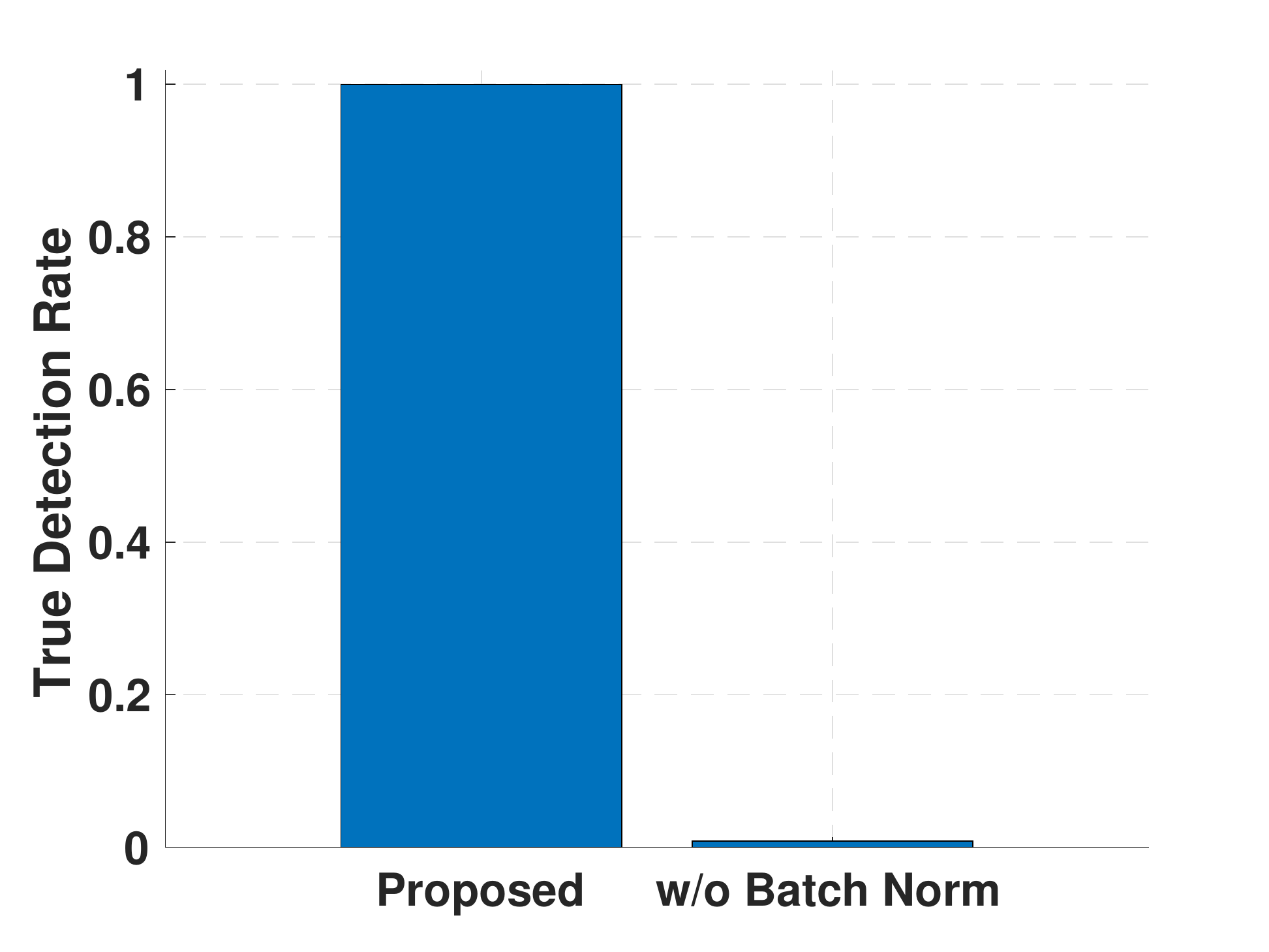}
        \label{fig:batchnorm_D3}
    }
    \subfigure[D4]{
         \includegraphics[scale = 0.3]{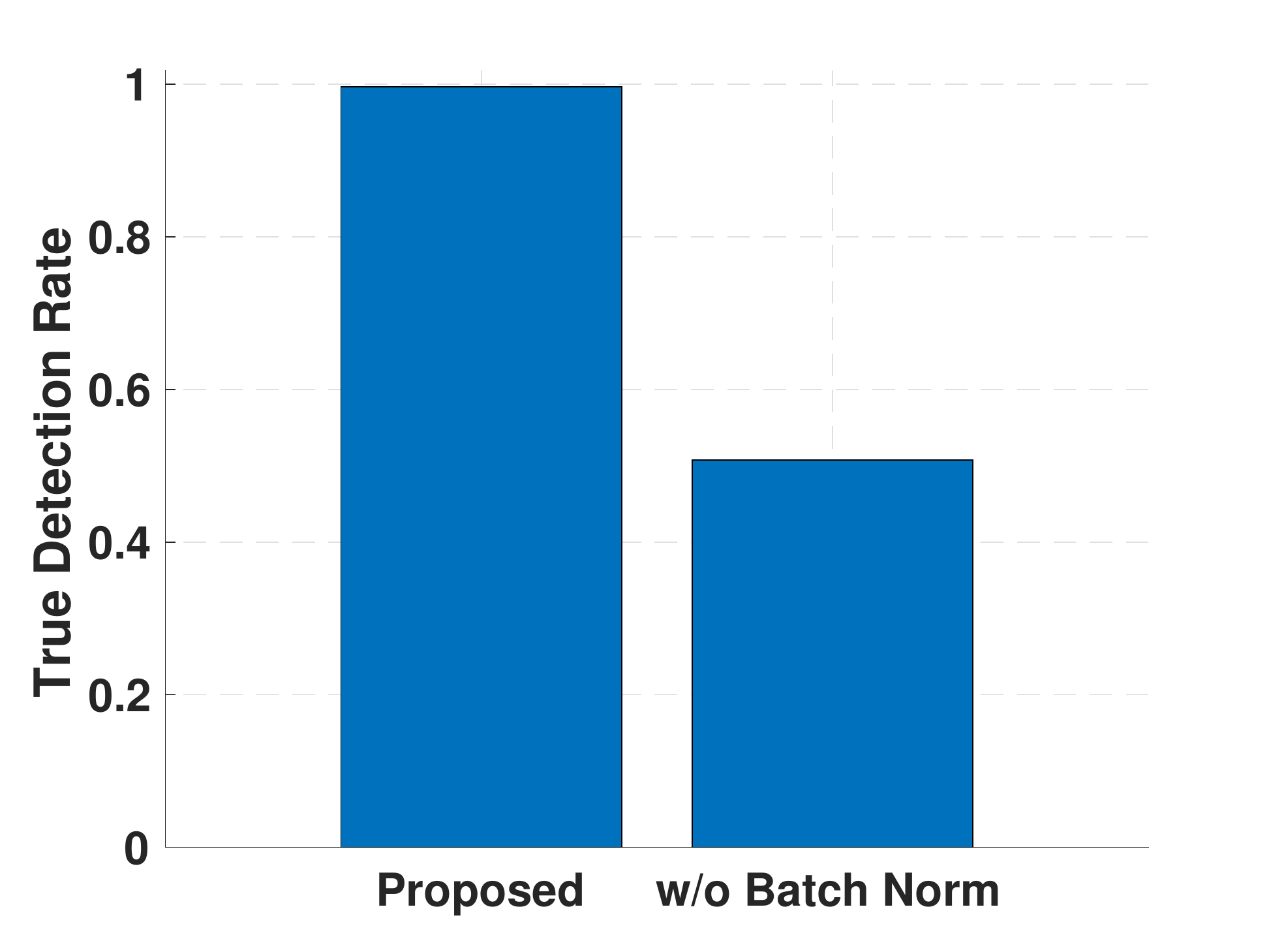}
        \label{fig:batchnorm_D4}
    }
    \caption{The true detection rate of the proposed deep learning model with/without the batch normalisation layer.}
   \label{fig:batchnorm}
\end{figure}

Because of the changing received signal strength, it is impossible to learn a feasible model without batch normalisation layer.

Figure \ref{fig:batchnorm} depicts the true detection rates achieved by applying the proposed DNN model with and without the batch normalisation layer. 
In D2 and D3, the true detection rates are no more than 2\% without the batch normalisation layer. 
Moreover, the performance of our model without the batch normalisation layer achieves 20.51\% and 50.75\% in D1 and D4, respectively. These results again demonstrate the importance of the batch normalisation layer in the DNN model for CSI based context awareness applications. 

\subsubsection{Effect of The Average Pooling Layer}
\begin{figure}[!ht]
    \centering
    \subfigure[D1]{
         \includegraphics[scale = 0.33]{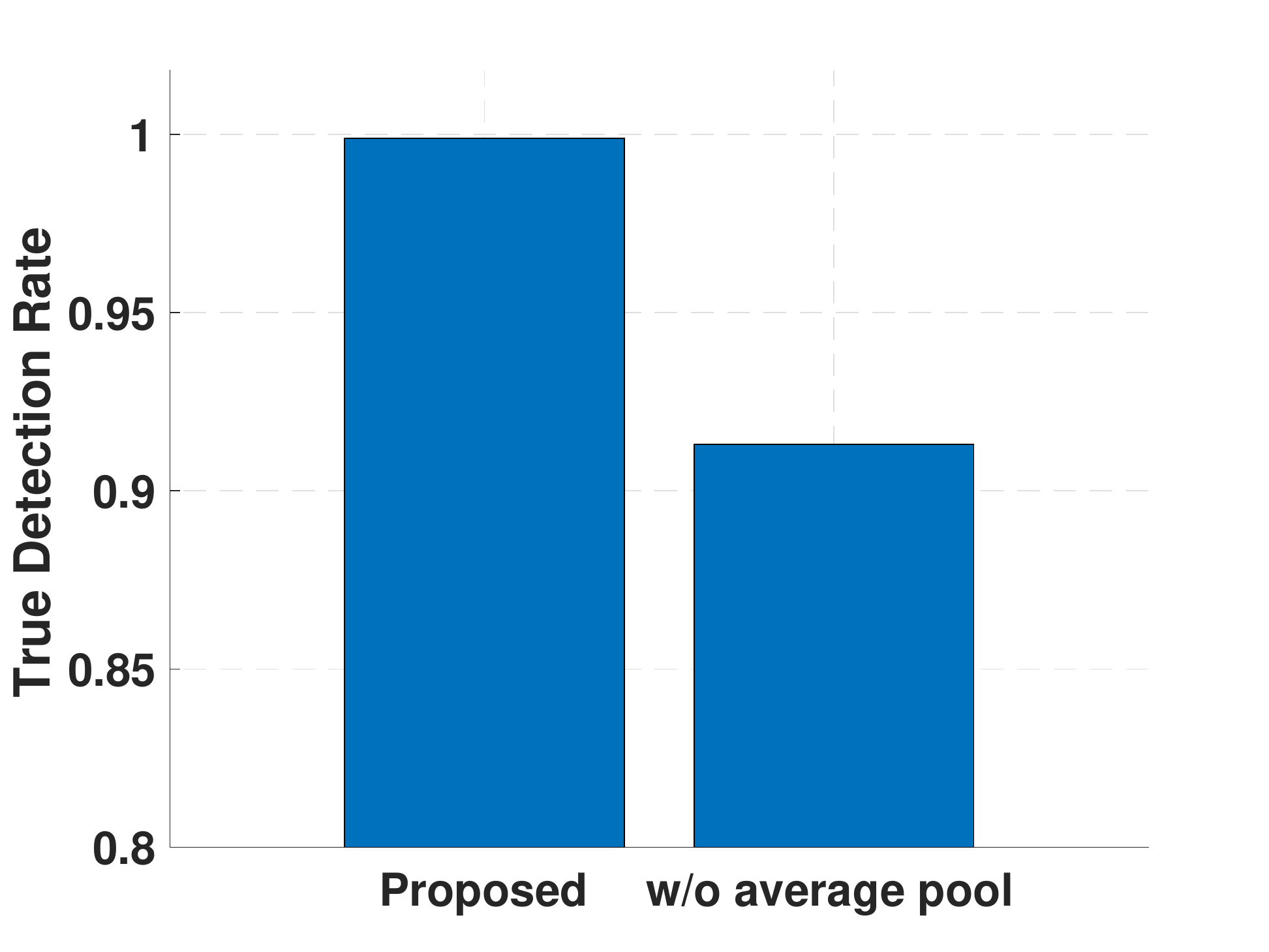}
        \label{fig:average_D1}
    }
    \subfigure[D2]{
         \includegraphics[scale = 0.33]{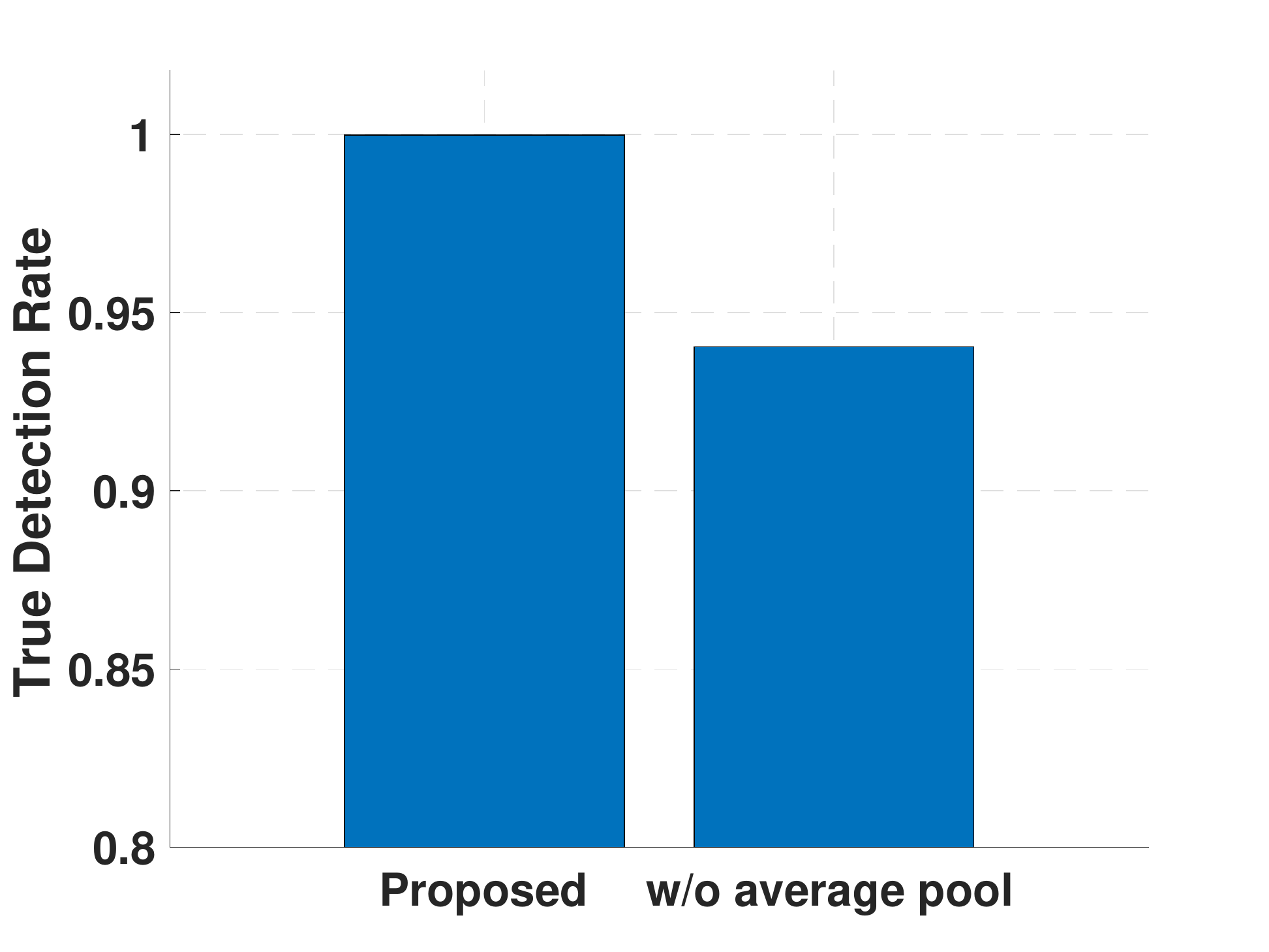}
        \label{fig:average_D2}
    }
    
    \subfigure[D3]{
         \includegraphics[scale = 0.33]{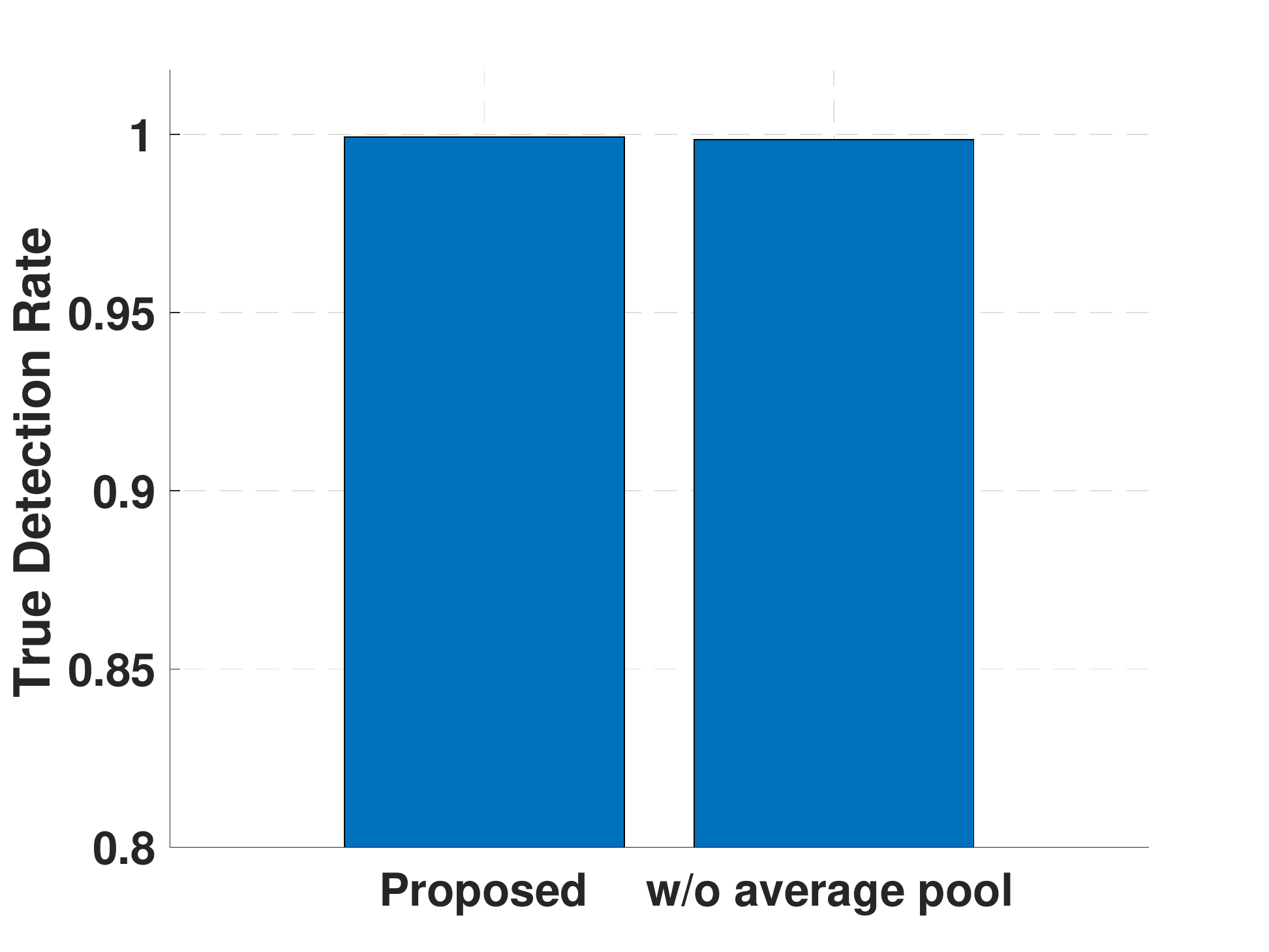}
        \label{fig:average_D3}
    }
    \subfigure[D4]{
         \includegraphics[scale = 0.33]{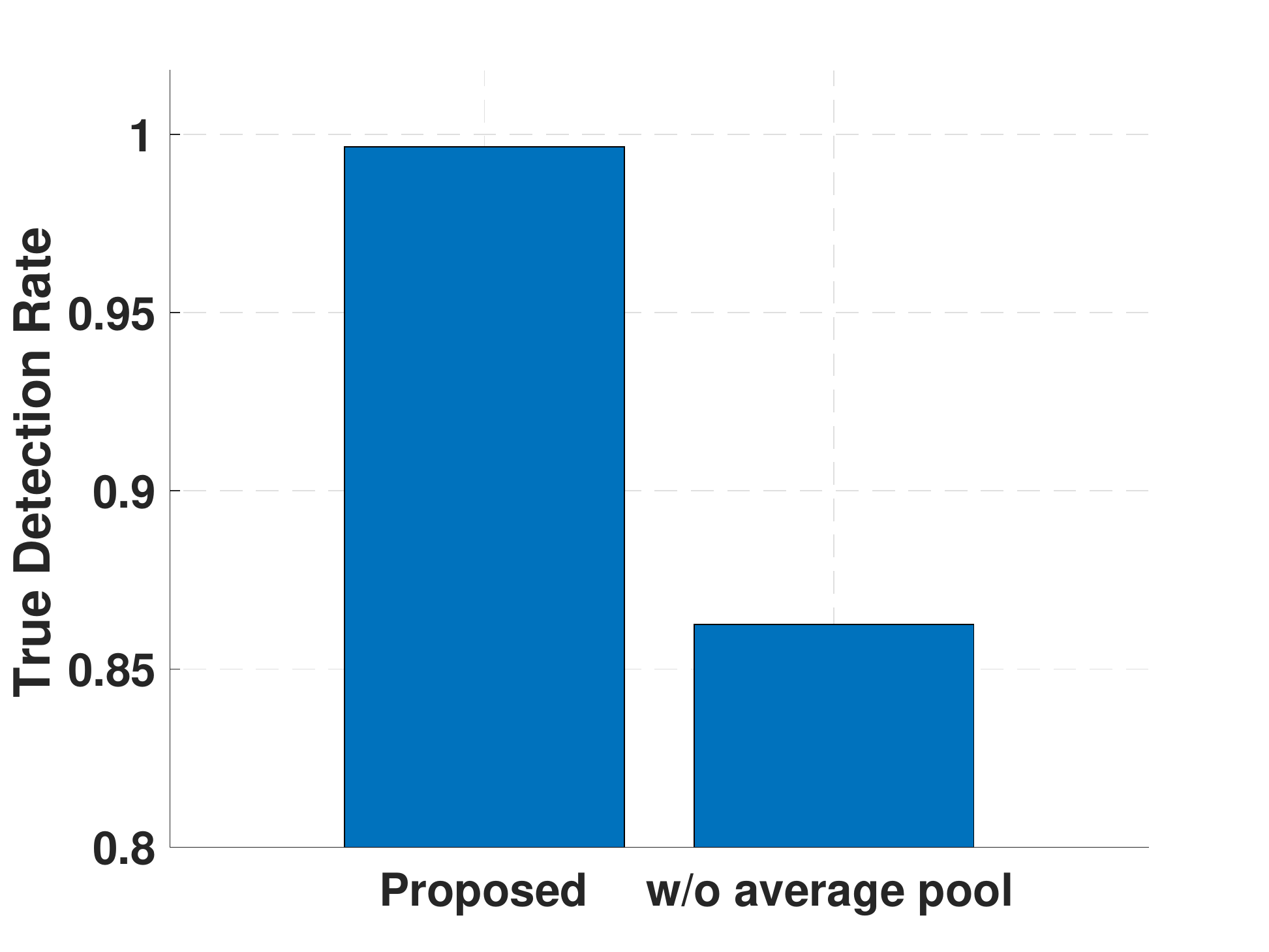}
        \label{fig:average_D4}
    }
    \caption{The effect of average pooling}
   \label{fig:average}
\end{figure}

Figure \ref{fig:average} shows the true detection rates of our proposed model with and without the average pooling layer. In the D1, D2 and D4, the true detection rates drop by 13.64\%, 5.94\% and 13.4\% to 86.25\%, 94.04\% and 86.25\% respectively, and the performances are similar in D3. 
These results show that average pooling layers play a vital role in the performance stability of the radio-based context awareness.


\subsection{The Performance in Activity Dataset} \label{subsec:exp_activity}
\subsubsection{Dataset Description}
The Activity dataset is from the research work \cite{wei2019from}. In \cite{wei2019from}, authors use software-defined radio WAPS nodes to simulate various 802.11 protocols for radio-based device-free activity recognition. Two WASP nodes were deployed in a one-bedroom apartment. One WASP node was the transmitter, which sent one packet every 0.1 second. The other WASP node, as the receiver, received packets and transferred CSI data to a connected PC through a USB cable for further processing.  

Two primary datasets have been collected in two radio frequency conditions, i.e. with and without radio frequency interference (RFI). When collecting the CSI data, WAPS operated in 5.8GHz and used the entire 125 MHz industrial, scientific and medical (ISM) radio band. 8 common location-oriented activities were investigated in \cite{wei2019from}. In the RFI environment, one router was deployed in the middle of the apartment and communicated with the PC to generate radio interference in Channel 157 whose bandwidth is 20 MHz. Two secondary datasets in Channel 157, with and without RFI, were selected to explore the performance of the radio-based device-free activity recognition in different RFI conditions. 
To evaluate our proposed general framework, we use these two secondary datasets, denoted as AD1 (Channel 157 without RFI) and AD2 (Channel 157 with RFI). Each dataset has 952 instances. In each instance, there are 5 samples, 52 subcarriers and 1 antenna pair. 
We also use the same setting 10-fold cross-validation in \cite{wei2019from} to evaluate our proposed framework. 


\begin{table}[]
\caption[Table caption text]{Configuration of DNN used in Activity datasets}\label{tab:con_dnn_act}
\centering
\begin{tabular}{lcc}
\hline
                & Kernel Size                & Stride               \\ \hline
Conv\_1         & 2 x 1                      & 2 x 1                \\
Conv\_2         & 1 x 2                      & 1 x 1                \\
Conv\_3         & 1 x 3                      & 1 x 1                \\
Conv\_4         & 1 x 4                      & 1 x 1                \\
Conv\_5         & 1x 8                       & 1 x 1                \\
Conv\_6         & 1 x 12                     & 1 x 1                \\
Conv\_7         & 1 x 16                     & 1 x 1                \\ \hline \hline
                & \multicolumn{2}{c}{Pool Size}                     \\ \hline
AP\_1           & \multicolumn{2}{c}{1 x 2}                         \\
AP\_2           & \multicolumn{2}{c}{1 x 3}                         \\
AP\_3           & \multicolumn{2}{c}{1 x 4}                         \\ \hline \hline
                & \multicolumn{2}{c}{Number of Unit / Dropout Rate} \\ \hline
Fully connected & \multicolumn{2}{c}{1000}                          \\
Dropout         & \multicolumn{2}{c}{0.8}                           \\
Fully connected & \multicolumn{2}{c}{1000}                          \\
Dropout         & \multicolumn{2}{c}{0.8}                           \\ \hline
\end{tabular}
\end{table}

\subsubsection{Architecture of the DNN model for Activity Dataset}
Table \ref{tab:con_dnn_act} shows the architecture of the DNN model we use for AD1 and AD2. This architecture still uses our proposed general framework but different settings with SignFi dataset. The reasons for the different settings are two folds. First, the inputs of the CSI data are different. There is only one transmit-receive antenna pair used in the configuration of the Activity dataset, but CSI from multiple antenna pairs are collected in SignFi datasets. More subcarriers from one channel can be accessed to provide CSI from WASP device( i.e. 52 subcarriers instead of 30). Secondly, the Activity datasets consider the RFI in one secondary dataset AD2. RFI significantly increases the difficulties of pattern analysis, which is required to include additional efforts in the DNN model design. 

Table \ref{tab:con_dnn_act} shows the model used for the Activity datasets.
In this model, we employ 7 convolutional neural network layers, 3 parallel average pooling layers followed with a concatenate layer, and 2 fully connected layers each followed with a dropout layer. As discussed in the general framework, the first convolutional neural network layer considers the complex-valued input with $2 \times 1$ kernel size and stride. More convolutional neural network layers are added in this model to address the challenges of noisy data and compensate for the lack of available antenna pairs. 3 average pooling layers are attached with pool sizes $1 \times 2$, $1 \times 3$ and $1 \times 4$. We use less average pooling layers than that used on the SignFi datasets because the usable patterns in the neighbouring samples decrease with the impact of the RFI. Furthermore, the consecutive samples in the AD2 with RFI may not be correlated, so we set the first dimension of the pool sizes of these three average pooling layers as 1. In this case, we only consider one sample in each average pooling layer. To further improve the performance, we use 1 more fully connected layer. The softmax layer is used as the final layer for training and inference.


\begin{figure}[!ht]
    \centering
    \subfigure[AD1]{
         \includegraphics[scale = 0.33]{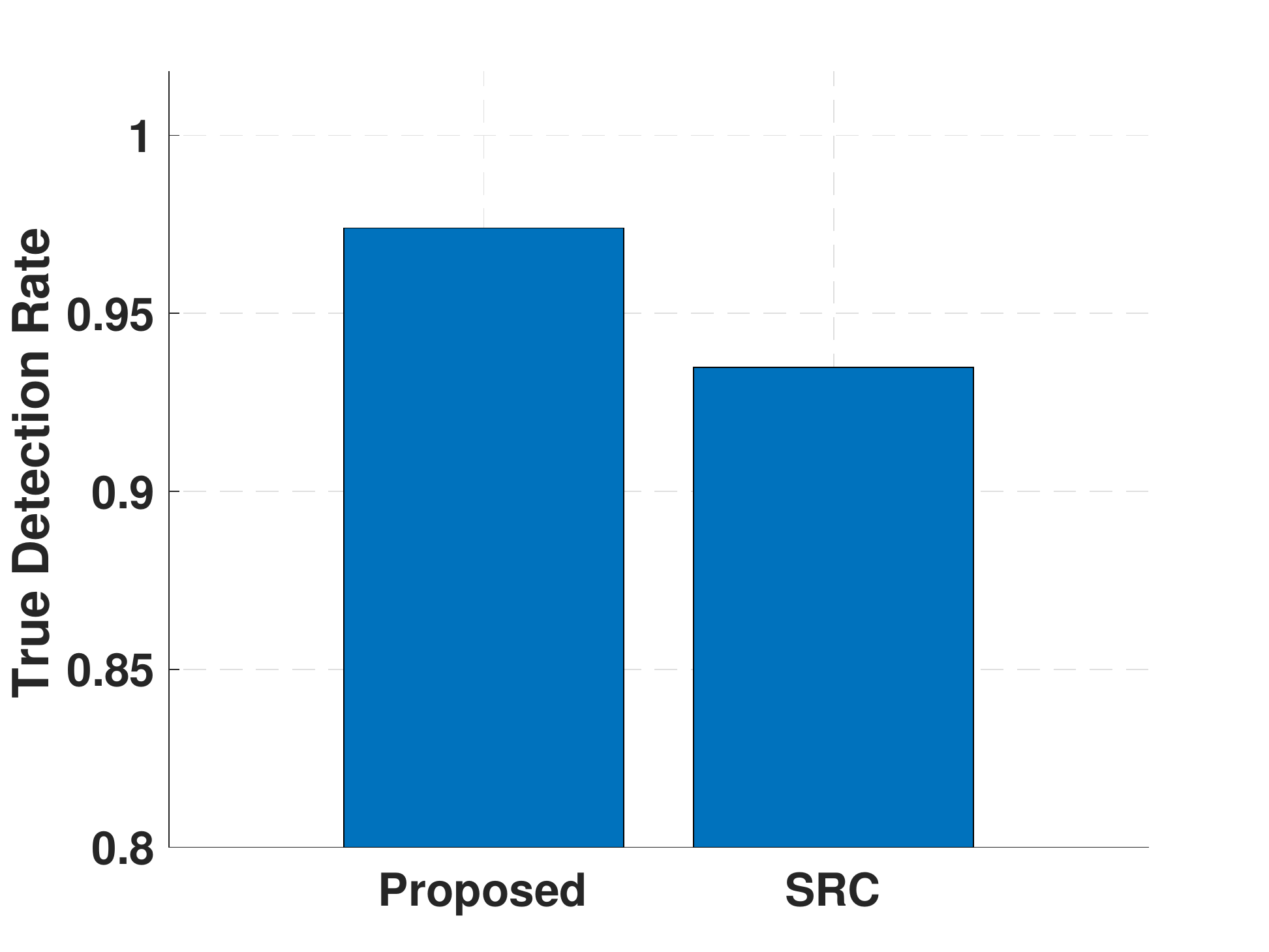}
        \label{fig:wasp_performance_D1}
    }
    \subfigure[AD2]{
         \includegraphics[scale = 0.33]{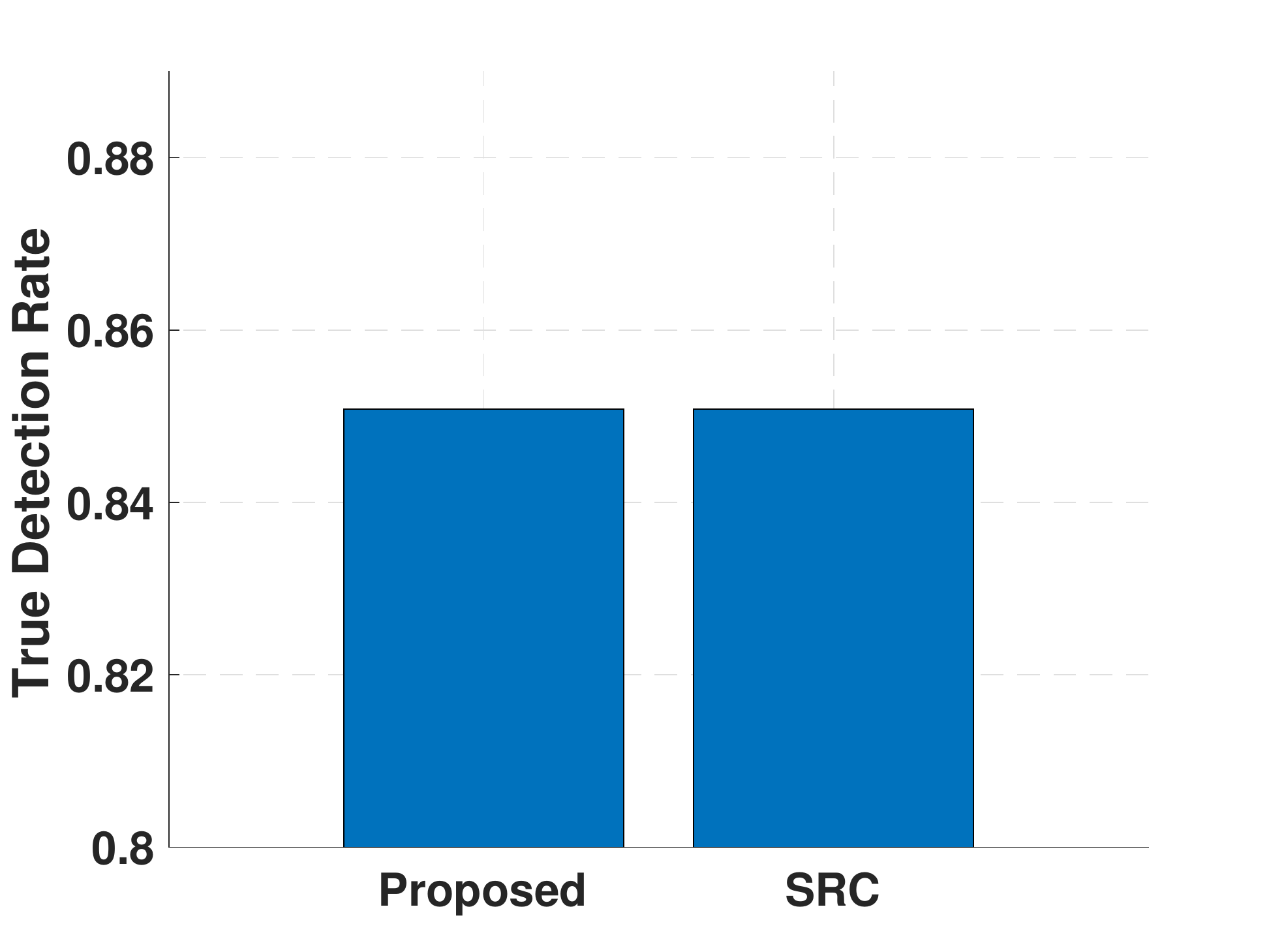}
        \label{fig:wasp_performance_D5}
    }
    \caption{The performance of our proposed architecture and the method in \cite{wei2015radio}}
   \label{fig:wasp_performance}
\end{figure}

\subsubsection{Comparing with the benchmark}
In this section, we compare the performance using our proposed general framework with the method in the previous research work \cite{wei2019from}. Figure \ref{fig:wasp_performance} shows the accuracy of our proposed architecture and the method in \cite{wei2015radio}. The true detection rate increases from 93.48\% to 97.40\% in AD1 without RFI, which indicates the advantages of our proposed deep learning based architecture. The true detection rate using our proposed framework in AD2 is 85.08\% the same as that in \cite{wei2019from}. Please note our general framework do not need any signal processing, while the previous method requires the explicit complex-valued CSI processing method design \cite{wei2015radio,sen2012you}. The results confirm the feasibility of the application of our proposed deep learning based architecture in challenging RFI environments.

\begin{figure}[!ht]
    \centering
    \subfigure[AD1]{
         \includegraphics[scale = 0.33]{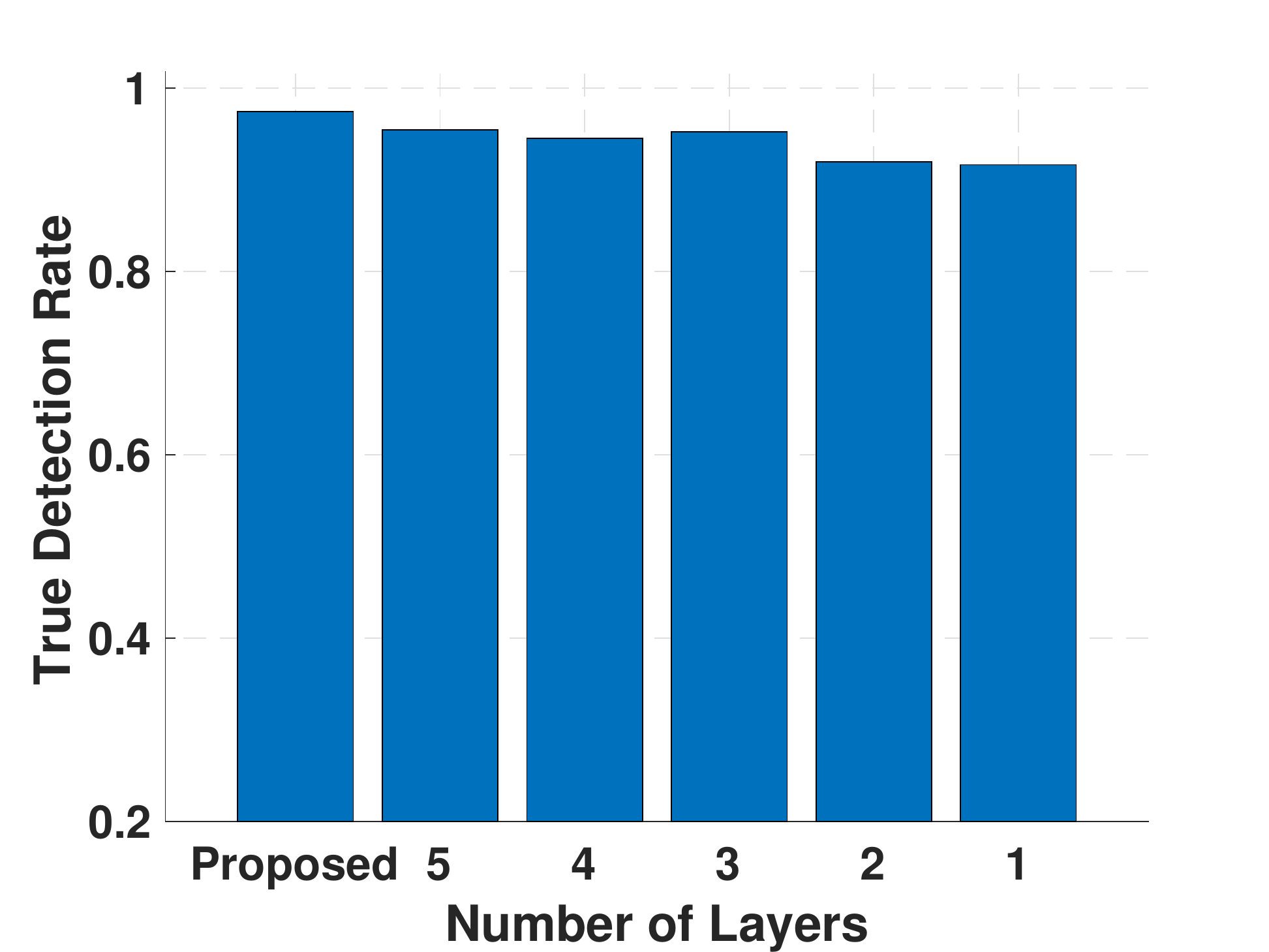}
        \label{fig:wasp_conv_d1}
    }
    \subfigure[AD2]{
         \includegraphics[scale = 0.33]{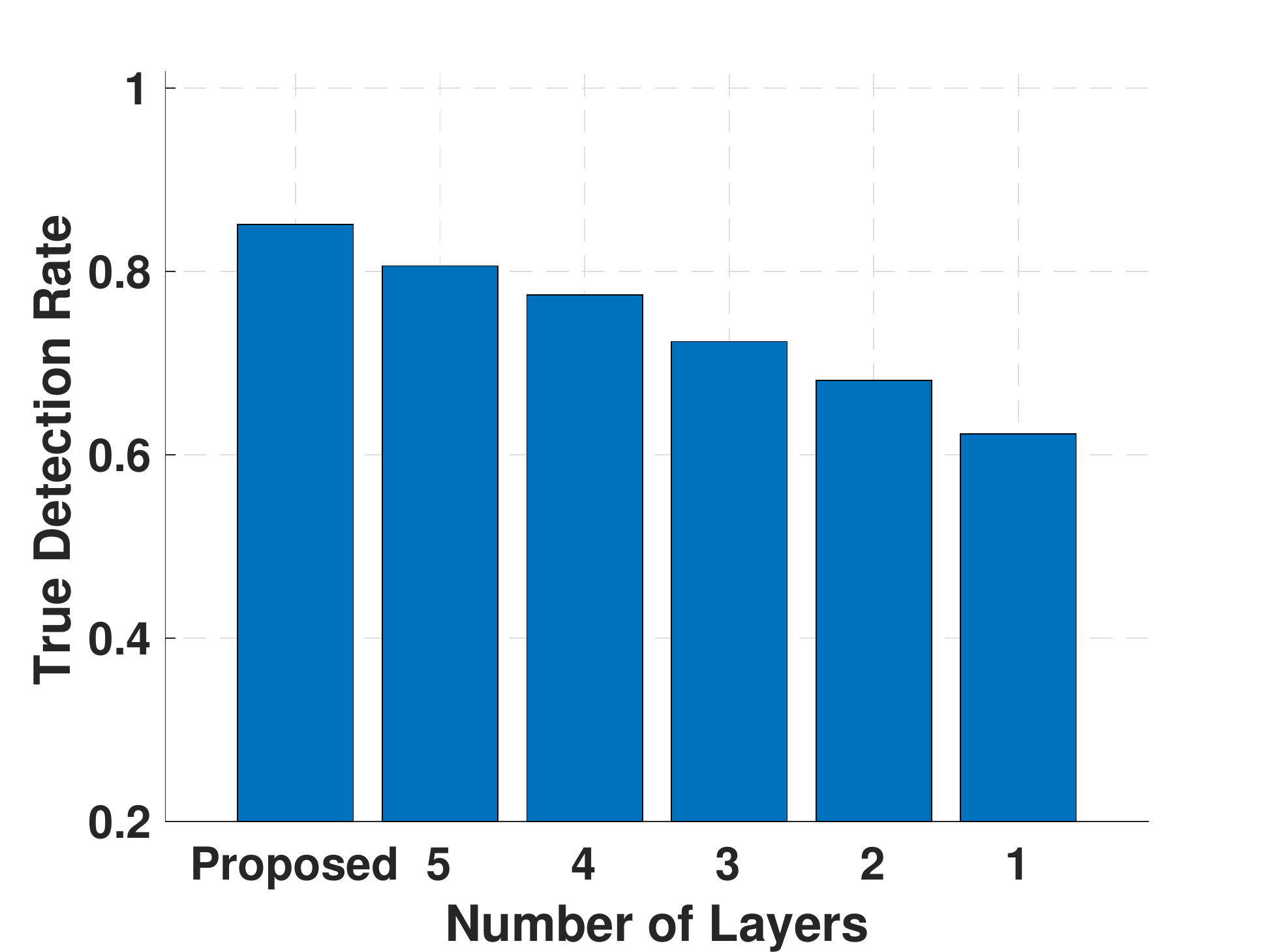}
        \label{fig:wasp_conv_d5}
    }

    \caption{The true detection rate of the proposed deep learning model with different number of Convolutional Layers}
   \label{fig:wasp_conv}
\end{figure}

\subsubsection{Effect of the Convolutional Layer}
In the section, we show the effect of convolutional layers in Activity datasets. Because Conv\_1 layer is designed for the complex-valued CSI, we only explore the effect of the remaining 6 layers in this section. Figure \ref{fig:wasp_conv} shows the performance of our proposed architecture with different numbers of convolutional layers. Figure \ref{fig:wasp_conv_d1} demonstrates that the number of convolutional layers does not impact to the performance in AD1 without RFI. The similar results are also observed in SignFi D3 shown in Figure \ref{fig:conv_D3}. The CSI data without RFI usually contain distinct patterns which can be easily explored for context awareness without considerable effort. However, when introducing RFI, radio-based context awareness becomes much more challenging. There is an apparent trend of the accuracy decrease with the reduction of the number of convolutional layers, as shown in Figure \ref{fig:wasp_conv_d5}. This confirms that the RFI introduces the extra challenges to the radio-based context awareness applications, and incorporating more convolutional layers can increase its performance of deep learning based context awareness model.

\begin{figure}[!ht]
    \centering
    \subfigure[AD1]{
         \includegraphics[scale = 0.33]{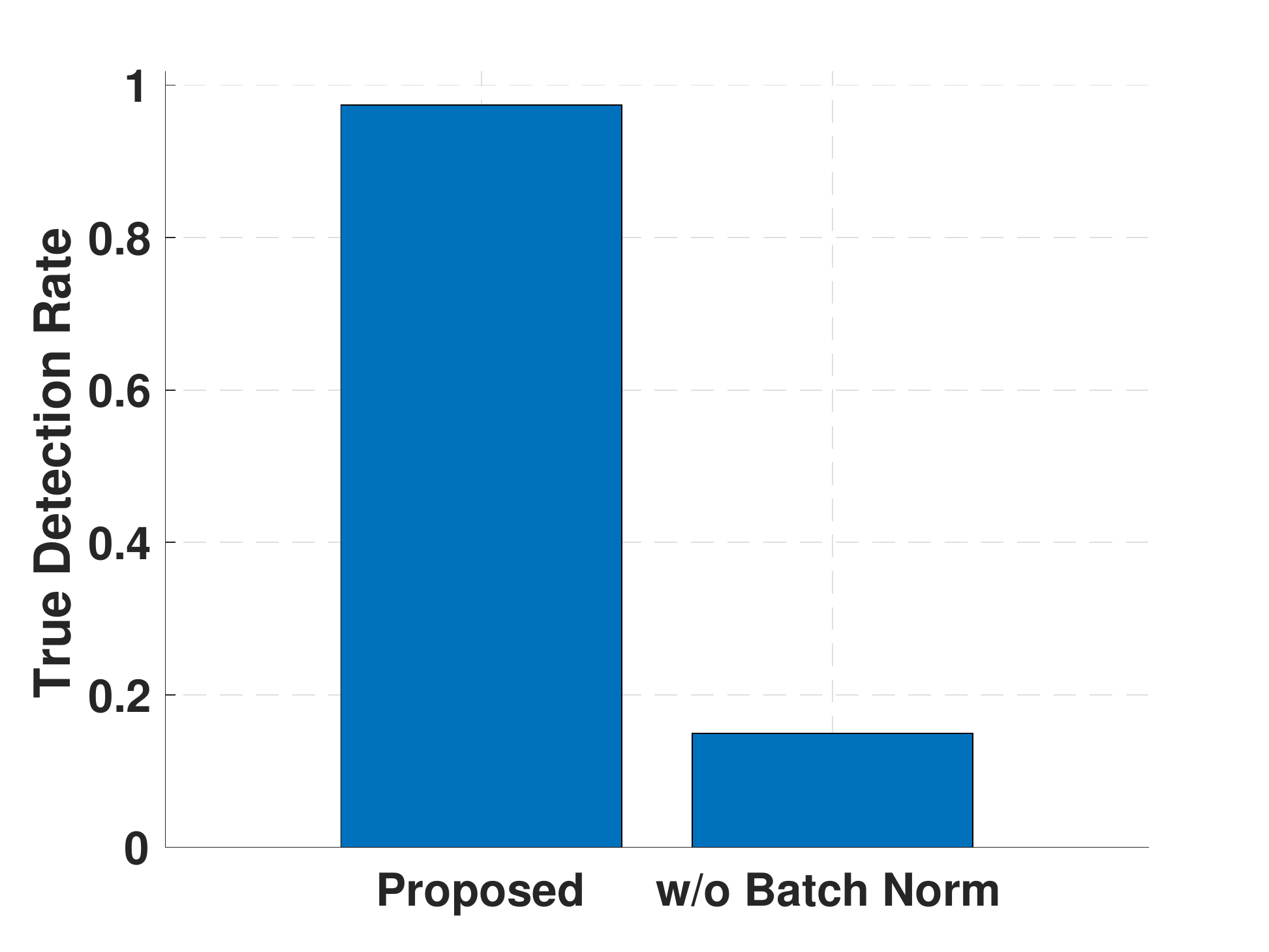}
        \label{fig:wasp_batchnorm_D1}
    }
    \subfigure[AD2]{
         \includegraphics[scale = 0.33]{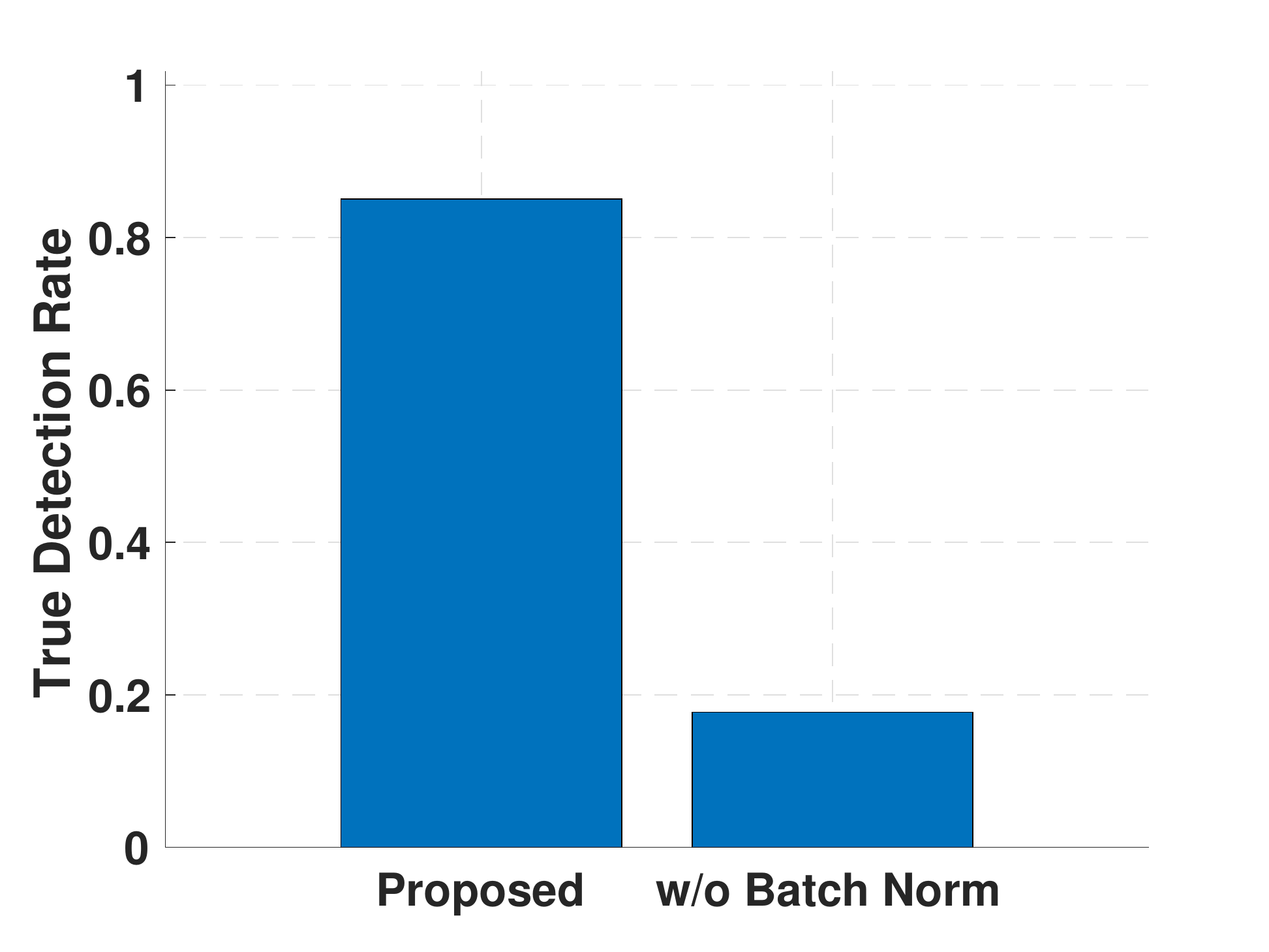}
        \label{fig:wasp_batchnorm_D2}
    }
    \caption{The true detection rate of the proposed deep learning model with and without batch normalisation layers.}
   \label{fig:wasp_batchnorm}
\end{figure}

\subsubsection{Effect of The Batch Normalisation Layer}

It is observed from Section \ref{sec:exp_norm} that Normalisation can significantly improve inference performance and accelerate the training process. This is further confirmed in the Activity datasets that the batch normalisation layers can help analyse patterns in radio-based device-free context awareness applications. As shown in Figure \ref{fig:wasp_batchnorm}, the accuracy is below 20\% without using batch normalisation layers in both AD1 and AD2. When applying the batch normalisation layers along with convolutional layers, it is conspicuous that the accuracy increases more than 80\% and 60\% in AD1 and AD2, respectively.

\begin{figure}[!ht]
    \centering
    \subfigure[AD1]{
         \includegraphics[scale = 0.33]{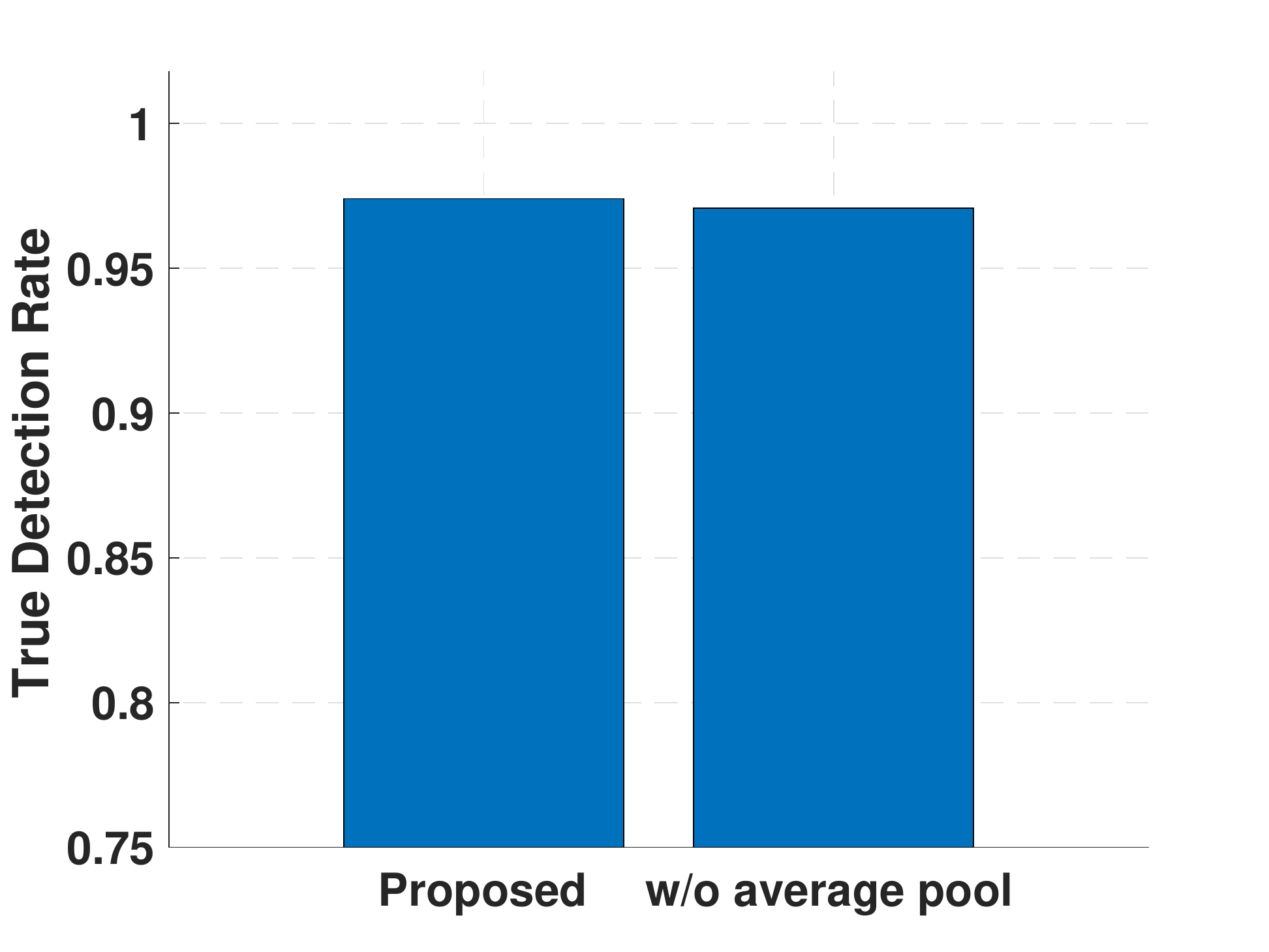}
        \label{fig:wasp_average_D1}
    }
    \subfigure[AD2]{
         \includegraphics[scale = 0.33]{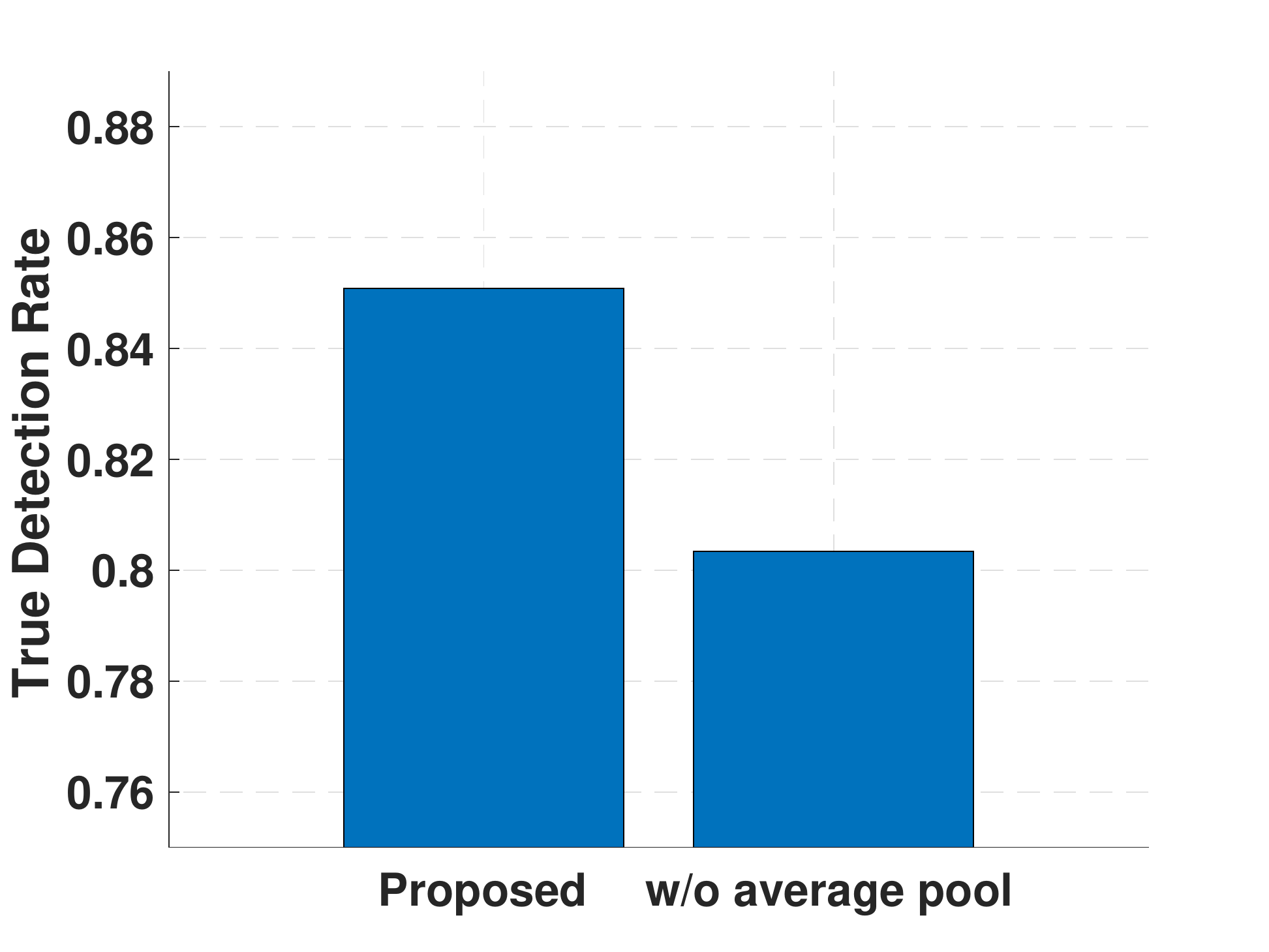}
        \label{fig:wasp_average_D2}
    }
    \caption{The true detection rate of the proposed deep learning model with and without average pooling}
   \label{fig:wasp_average}
\end{figure}

\subsubsection{Effect of The Average Pooling Layer}
Finally, we will discuss the effect of average pooling layers. Figure \ref{fig:wasp_average} shows the true detection rates of activity recognition in AD1 and AD2 with and without average pooling layers. What can be clearly shown in Figure \ref{fig:wasp_average_D2} is the dramatic decrease of the accuracy from 85.08\% to 80.34\% in AD2 without average pooling layers, while Figure \ref{fig:wasp_average_D1} shows the accuracy remains steady in AD1 without RFI. This reveals that average pooling layers can improve the performance in RFI environment.